\documentclass[12pt,preprint]{aastex}

\usepackage{lscape,graphicx,rotating,amsmath,color}
\usepackage{multirow}

\definecolor{orange}{cmyk}{0,0.4,0.8,0.2}
\definecolor{darkorange}{rgb}{.71,0.21,0.01}
\definecolor{darkgreen}{rgb}{.12,.54,.11}

\usepackage{hyperref}
\hypersetup{
  breaklinks=true,  
  colorlinks=true,
  urlcolor=blue,
  linkcolor=darkorange,
  citecolor=darkgreen,
  }
\usepackage{breakurl}

\newcommand{\hip}{\textit{Hipparcos}}
\begin{document}

\title{Optimizing Automated Classification of Periodic Variable Stars in New Synoptic Surveys}

\author{James P. Long\altaffilmark{1}, Noureddine El Karoui\altaffilmark{1}, John A. Rice\altaffilmark{1}, Joseph W. Richards\altaffilmark{1,2}, and Joshua S. Bloom\altaffilmark{2}}

\altaffiltext{1}{Department of Statistics, University of
  California, Berkeley, CA 94720-3860, USA.}

\altaffiltext{2}{Astronomy Department, University of
  California, Berkeley, CA 94720-7450, USA.}


\shorttitle{Optimizing Automated Classification of Periodic Variable Stars from New Surveys}
\shortauthors{Long \textit{et al.}}


\begin{abstract}
Efficient and automated classification of periodic variable stars is becoming increasingly important as the scale of astronomical surveys grows. Several recent papers have used methods from machine learning and statistics to construct classifiers on databases of labeled, multi--epoch sources with the intention of using these classifiers to automatically infer the classes of unlabeled sources from new surveys. However, the same source observed with two different synoptic surveys will generally yield different derived metrics (features) from the light curve. Since such features are used in classifiers, this survey-dependent mismatch in feature space will typically lead to degraded classifier performance. In this paper we show how and why feature distributions change using OGLE and \hip{} light curves. To overcome survey systematics, we apply a method, \textit{noisification}, which attempts to empirically match distributions of features between the labeled sources used to construct the classifier and the unlabeled sources we wish to classify. Results from simulated and real--world light curves show that noisification can significantly improve classifier performance. In a three--class problem using light curves from \textit{Hipparcos} and OGLE, noisification reduces the classifier error rate from 27.0\% to 7.0\%. We recommend that noisification be used for upcoming surveys such as Gaia and LSST and describe some of the promises and challenges of applying noisification to these surveys. 
\end{abstract}

\keywords{classification, periodic variables, machine learning, statistics, errors in variables}

\section{Introduction}
\label{sec:intro}
Classification of periodic variables is crucial for scientific knowledge discovery and efficient use of telescopic resources for source follow up \citep{eyer2008variable,walkowicz2009impact}.
As the size of synoptic surveys has grown, a greater and greater share of the classification process must become automated \citep{bloom2011data}. With \hip{}, it was possible for astronomers to individually analyze and classify each of the 2712 periodic variables observed in the survey. Starting in 2013, Gaia is expected to discover $\sim$ 5 million classical periodic variables over the course of its 4--5-year mission \citep{eyer2000predictions}. LSST, for that matter, may collect on the order of a billion \citep{borne2007data}. Individual analysis and classification by hand of all periodic variables is no longer feasible.

The need for efficient and accurate source classification has motivated much recent work on applying statistical and machine learning methods to variable star data sets (e.g., \citealt{eyer2005automated,debosscher2007automated,richards2011machine,dubath2011random}). In these papers, classifiers were constructed  using light curves from a variety of surveys, such as the Optical Gravitational Lensing Experiment (OGLE, \citealt{2011AcA....61....1S}), \hip{} \citep{1997perr}, The All-Sky Automated Survey (ASAS, \citealt{acvs}), the COnvection, ROtation
\& planetary Transits survey (CoRoT, \citealt{2009A&A...506..411A}), and the Geneva Extrasolar Planet Search. Often the intention of these studies is to develop classifiers with high accuracy in classifying sources from surveys other than those used to construct the classifier. For example, \citet{blomme2011tres} trained a classifier on a mixture of \textit{Hipparcos}, OGLE, and CoRoT sources and used it to classify sources from the Trans-atlantic Exoplanet Survey (TrES, \citealt{2009nsted.cat....6O}) Lyr1 field. \citet{dubath2011random} and \citet{2008AIPC.1082..257E} view their work on classification of \hip{} sources as a precursor to classification of yet--to--be collected Gaia light curves. Debosscher and collaborators trained a classifier on a mixture of OGLE and \hip{} sources in attempts to classify CoRoT sources \citep{debosscher2007automated,debosscher2008OGLE,debosscher2009automated}.

It is well known that systematic differences in cadence, observing region, flux noise, detection limits, and number of observed epochs per light curve exist among surveys. Even within surveys there is heterogeneity in these characteristics. Most statistical classifiers assume that the light curves of a known class used to construct the classifier, termed \textit{training data}, and the light curves of unknown class which we wish to classify, termed \textit{unlabeled data}, share the same characteristics. This is unlikely to be the case when training and unlabeled light curves come from different surveys, or when the best-quality light curves of sources from each class are used to classify poorly sampled light curves of unknown class from the same survey.

\begin{figure}[ht]
\epsscale{0.45000000000000001}
\begin{center}
\begin{tabular}{cc}
$\begin{array}{cc}
\multicolumn{1}{l}{\mbox{(a)}} & \multicolumn{1}{l}{\mbox{(b)}} \\ \\ \\ [-.35in]
\plotone{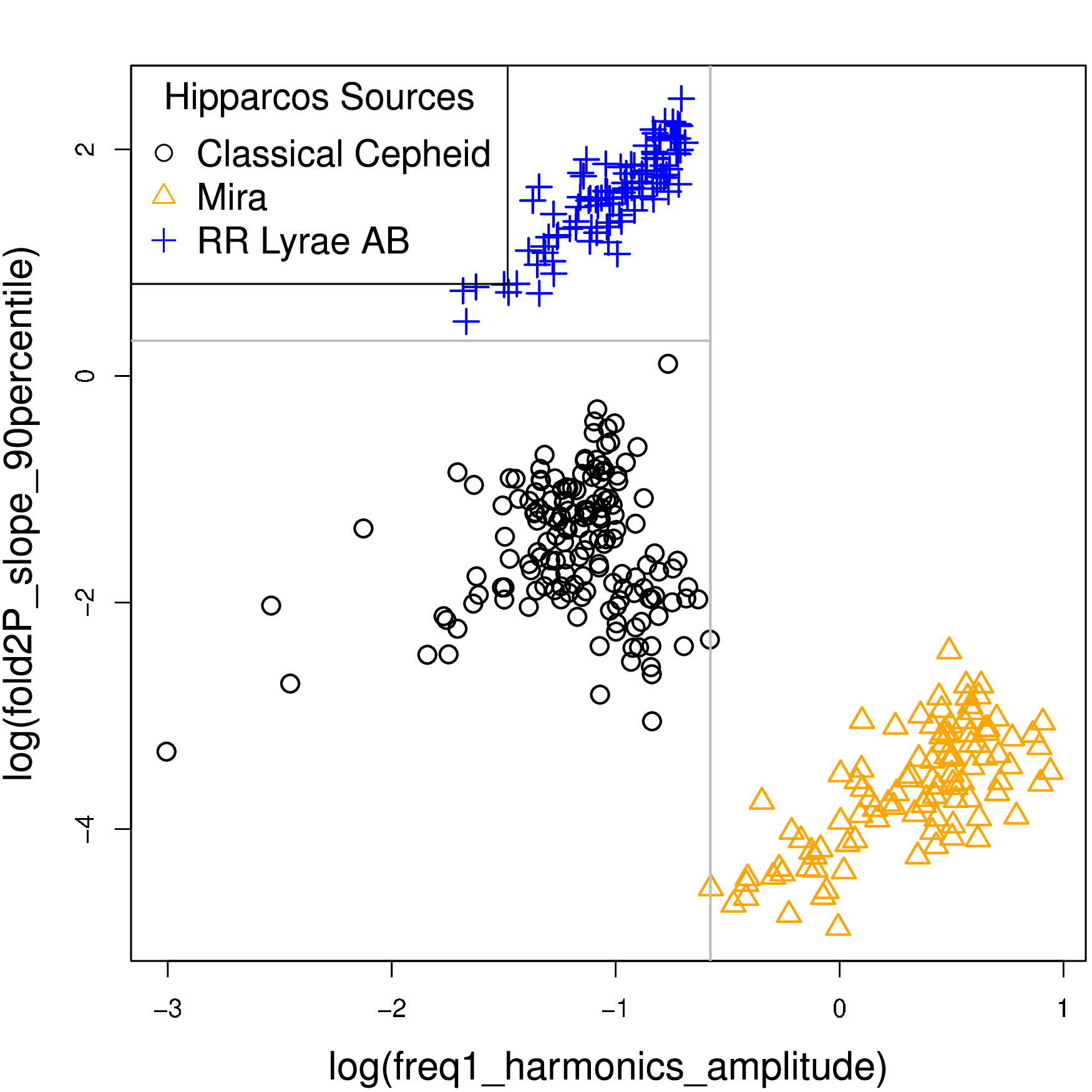} &
\plotone{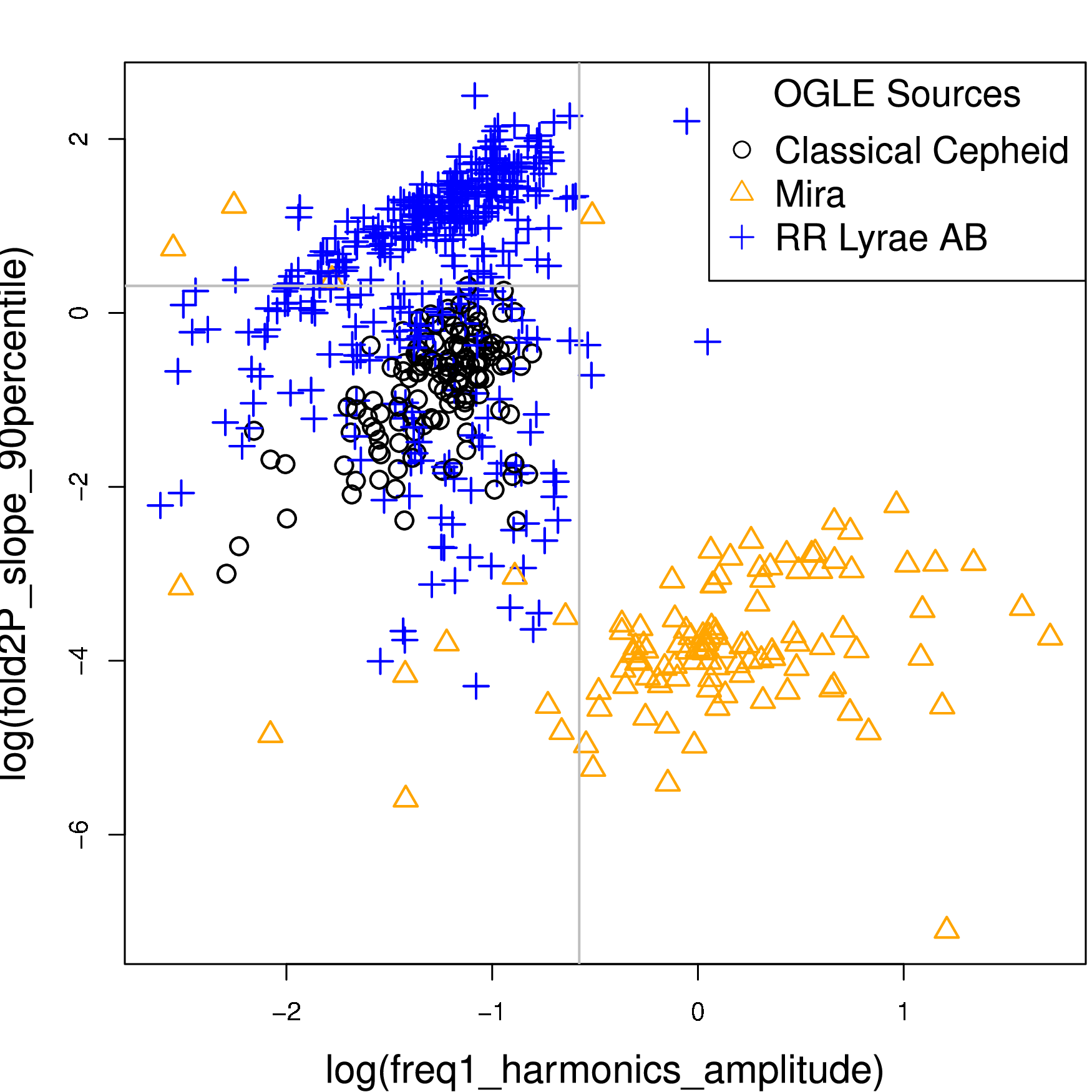}
\end{array}$
\end{tabular}
\end{center}
\caption{(a) The grey lines represent the CART classifier constructed using \hip{} data. The points are \hip{} sources. This classifier separates \hip{} sources well (0.6\% error as measured by cross--validation). (b) Here the OGLE sources are plotted over the same decision boundaries. There is now significant class overlap in the amplitude--fold2P plane (30\% error rate). This is due to shifts in feature distributions across surveys.}
\label{fig:cart_hip_ogle}
\end{figure}

To illustrate how seriously survey mismatches can deteriorate classification performance, consider the three-class problem of separating Mira variables, Classical Cepheids, and Fundamental Mode RR Lyrae from the \hip{} and OGLE surveys. From OGLE, we use V-band data. Note that OGLE is far better sampled in I-band than V-band. We use V-band to create a setting where one set of data is well sampled while the other set is poorly sampled. See Section \ref{sec:hip_to_classify_ogle} and Table \ref{tab:training_vs_unlabeled} for more information on these sources.

For each light curve we compute dozens of metrics, termed \textit{features}, that contain important information related to source class (e.g., frequency and amplitude; see Section \ref{sec:classification} for details on feature selection and extraction). Using the \hip{} light curves we construct a classifier using CART.\footnote{CART (Classification And Regression Trees) is a popular classifier that forms a sequence of nested binary partitions of feature space. See \cite{breiman1984classification} for more on CART.} The resulting classifier uses only two features for separating classes: the amplitude of a best fit sinusoidal model and the 90$^{th}$ percentile of the slope between phase adjacent flux measurements after the light curve has been folded on twice the estimated period.

Figure \ref{fig:cart_hip_ogle}a displays these two features for each \hip{} source with grey lines denoting the class boundaries chosen by CART. Based on the \hip{} light curves, this looks like an excellent classifier as each of the three regions of feature space selected by CART contains sources of only one class. However, examining a subset of the OGLE sources, Figure \ref{fig:cart_hip_ogle}b, shows large class overlap on these two features. Here these two features do not separate OGLE sources well. The error rate measured by cross--validation on the \hip{} sources was only 0.6\%\footnote{See \ref{sec:estimating_accuracy} for a definition of cross--validation}. However, the misclassification rate on the OGLE sources is 30.0\%.

Despite what the 30.0\% error rate seems to imply, the problem of separating classes in OGLE is not inherently difficult. A CART classifier trained on the OGLE light curves has a cross--validated error rate of 1.3\%. While there are many systematic differences between the \hip{} and OGLE surveys, their radically different cadences and number of flux measurements per light curve appear to be driving the increase in misclassification rate. For example, both features in Figure \ref{fig:cart_hip_ogle} depend on the estimate of each source's period; yet, over 25\% of the RR Lyrae in OGLE have incorrectly estimated periods due to poor sampling in the V-band.

A natural question to ask is: If we had observed the \hip{} sources at an OGLE cadence, what classifier would CART have constructed, and how would this have changed the error rate? In this paper we use \textit{noisification}, a method which matches the cadence of training data and unlabeled data by inferring a continuous periodic function for each training light curve and then extracting flux measurements at the cadence and photometric error level present in the unlabeled light curves. The purpose of noisification is to automatically shift the distribution of features in the training data closer to the distribution of features in the unlabeled data so that a classifier can determine class boundaries as they exist in the unlabeled data. Versions of noisification were introduced in \cite{starr2010map} and \cite{long2011classification}. In this paper, we demonstrate that noisification improves classification accuracy on several simulated and real--world data sets. For instance, on the OGLE -- Hipparcos three class problem we reduce misclassification rate by 20.0\%. Performance increases are greatest when the training data is well sampled at a particular cadence while unlabeled light curves are either poorly time sampled or observed at a different cadence.

This paper is organized as follows. In Section \ref{sec:classification} we briefly outline the statistical classification framework and show how it is applied in the context of periodic variables. In Section \ref{sec:feature_distributions} we illustrate the problems that occur when training and unlabeled data come from different surveys. We present \textit{noisification}, a method for overcoming differences related to number of flux measurements, cadence, and photometric error in Section \ref{sec:noisification}. In Section \ref{sec:data_experiments} we apply noisification to several data sets. Finally in Section \ref{sec:conclusions} we discuss possible uses of noisification for upcoming surveys.

\section{Overview of Classification of Periodic Variables}
\label{sec:classification}

Here we review a methodology for constructing, implementing, and evaluating statistical classifiers for periodic variables. This approach has been used in many recent works. For a more detailed review of the methodology see \cite{debosscher2007automated} or \cite{richards2011machine}.

\subsection{Constructing a Classifier}
We start with a set of light curves of known class, termed training data and a set of light curves of unknown class, termed unlabeled data. Our goal is to determine the classes for the unlabeled light curves using information present in the training data. Each light curve consists of a set of time, flux, and photometric error measurements. We compute functions of the time, flux, and photometric error, termed features. Features are chosen to contain information relevant for differentiating classes. The same set of features is computed for each light curve. A statistical classification method uses the training data to learn a relationship between features and class and produces a classifier $\mathcal{C}$. Given the features, $\mathbf{x}$, for a light curve in the unlabeled set, $\mathcal{C}(\mathbf{x})$ is a prediction of its class.

\subsection{Feature Set}
We use a total of 62 features to describe each light curve. 50 of these features are described in Tables 4 and 5 of \cite{richards2011machine}.\footnote{We do not use \textit{pair\_slope\_trend}, \textit{max\_slope}, or \textit{linear\_trend}.} We use 12 other features, described in Appendix \ref{app:new_features} of this article. Many of the features that we use are obvious choices e.g., frequency and amplitude. Most of our features, or features very similar to the ones here, have been used in recent work on classification of periodic variables \citep{kim2011qso,dubath2011random}.

\subsection{Choosing a Classifier}
There are many statistical classification methods for constructing the function $\mathcal{C}$. Some of the most popular include linear discriminant analysis (LDA), neural networks, support vector machines (SVMs), and Random Forests. In an earlier example we used CART. Each classification method has its own strengths and weaknesses. See \cite{hastie2009elements} for an extensive discussion of classification methods. In this work we use the Random Forests classifier developed by \cite{breiman2001random}, \cite{amit1997shape}, and \cite{dietterich2000experimental}. Random Forests has been used, with high levels of success, in recent studies of automated variable star classification \citep{richards2011machine,dubath2011random}. \cite{richards2011machine}, in a side--by--side comparison of 10 different classifiers using OGLE and \hip{} data, found that Random Forest had the lowest error rate.

\subsection{Estimating Classifier Accuracy}
\label{sec:estimating_accuracy}
Usually, researchers want an estimate of how accurate the classifier, $\mathcal{C}$, will be when presented with new, unlabeled data. Simply calculating the proportion of times $\mathcal{C}$ correctly classifies light curves in the training data is a poor estimate of classifier success, as this typically overestimates classifier performance on unlabeled data. Better assessment of classifier performance on unlabeled data is attained by using training--test set splits or cross--validation. With training--test set splits a fraction of the data, usually between 10\% and 30\%, is ``held out'' while the rest of the data is used to train the classifier. Subsequently, the held out observations are classified and the accuracy recorded. This number provides an estimate of how well the classifier will perform on unlabeled observations. In cross--validation, the training--test split is repeated many times, holding out a different set of observations at each iteration. The accuracy of the classifier is recorded at each iteration and then averaged. See Chapter 7 of \cite{hastie2009elements} for more information on assessing classifier performance. Cross--validation has been the method of choice for evaluating classifier performance in many of the recent articles on classification of periodic variables.

\section{Feature Distributions and Survey Systematics}
\label{sec:feature_distributions}

The classification framework described above comes with assumptions and limitations. Of critical importance, statistical classification methods are only designed to produce accurate classifiers when the relationship between features and classes is the same in training and unlabeled data. This is formalized as follows. Let $z$ represent the class for a source with features $\mathbf{x}$. Let $p_{tr}(z|\mathbf{x})$ be the probability of class given features in the training set and $p_{u}(z|\mathbf{x})$ be the probability of class given features for unlabeled data. Statistical classifiers are designed to have high accuracy when $p_{tr}(z|\mathbf{x}) = p_{u}(z|\mathbf{x}).$ In the three class example in the introduction, we saw that this was not the case due, in part, to incorrect estimation of periods in the unlabeled (OGLE) light curves. Violating this assumption will also cause cross--validation to make incorrect predictions of classifier accuracy.

In this section we illustrate the complex connection between survey systematics and feature distributions. We show how this connection causes the $p_{tr}(z|\mathbf{x}) = p_{u}(z|\mathbf{x})$ assumption to break, potentially leading to poor classifier performance on the unlabeled data.

\subsection{Periodic Features}

\begin{figure}[ht]
\epsscale{0.45000000000000001}
\begin{center}
\begin{tabular}{cc}
$\begin{array}{cc}
\multicolumn{1}{l}{\mbox{(a)}} & \multicolumn{1}{l}{\mbox{(b)}} \\ \\ \\ [-.35in]
\plotone{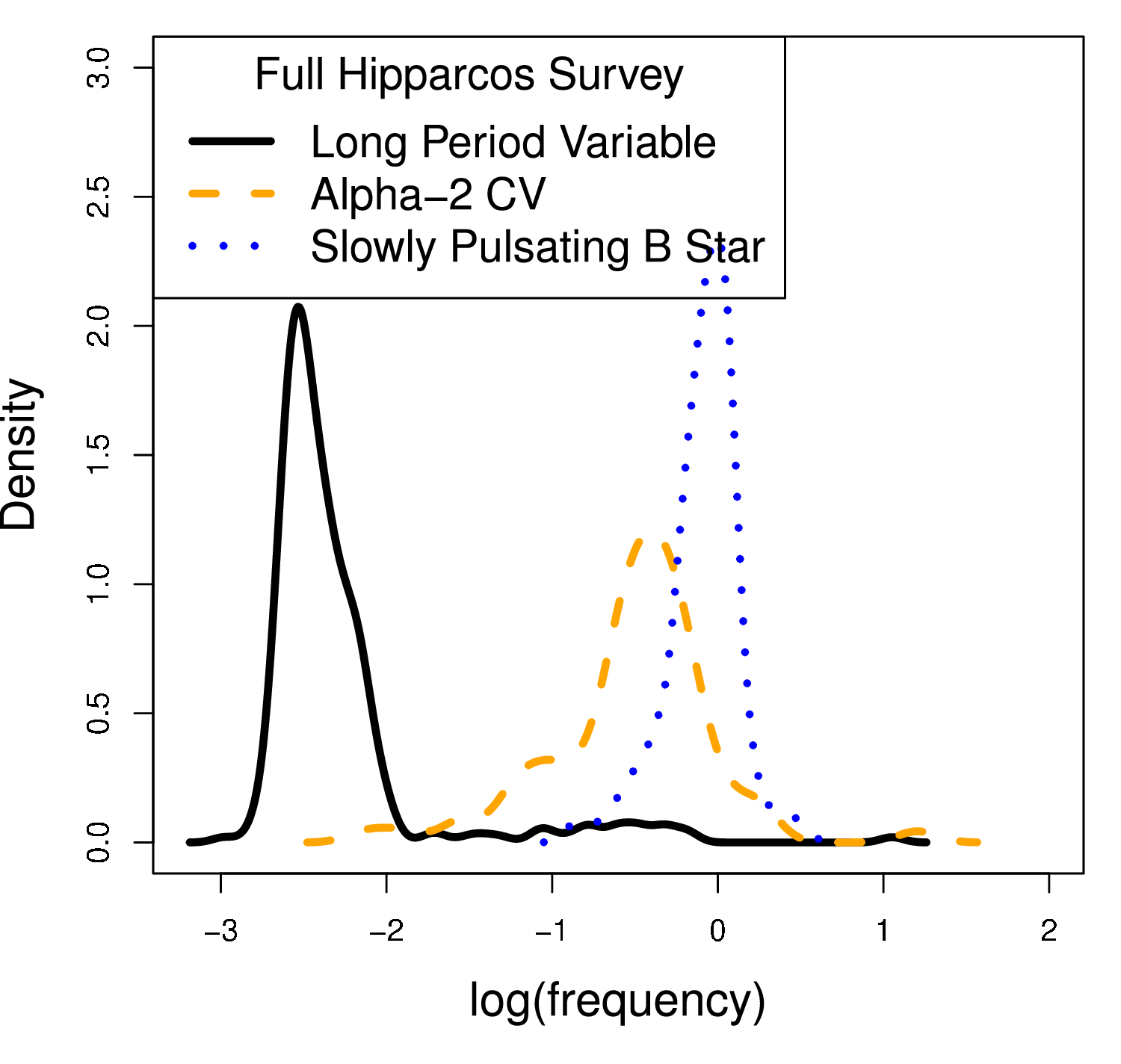} &
\plotone{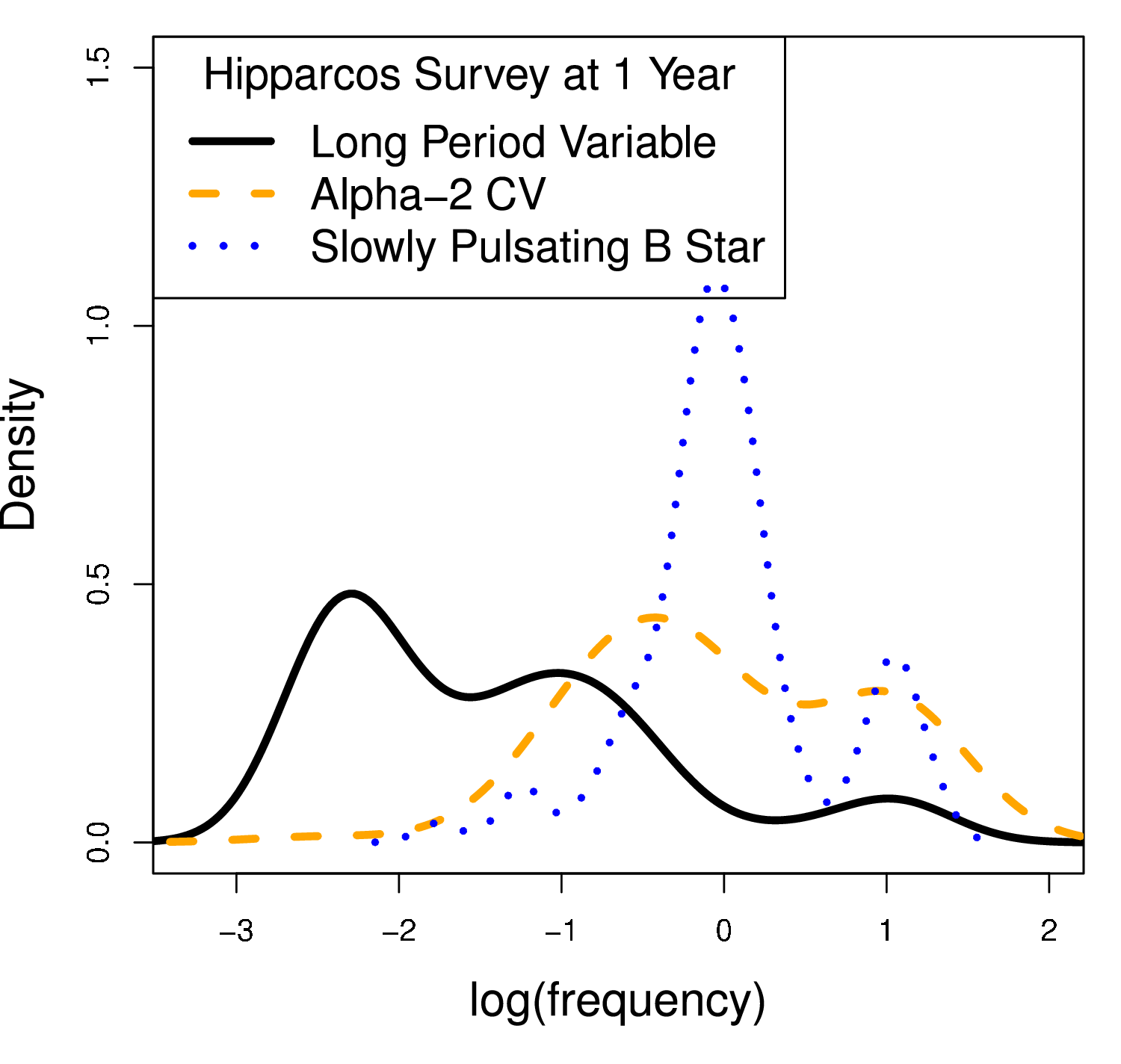}
\end{array}$
\end{tabular}
\end{center}
\caption{(a) Distribution of frequency for three source classes observed for entire length of \hip{}. (b) Distribution of frequency for same three sources classes observed for first 365 days of \hip{}. A classifier constructed on the the complete \hip{} light curves is likely to have poor performance on the \hip{} curves truncated to 365 days. This scenario could happen if \hip{} light curves were used to construct a classifier that was then applied to short light curves from the first Gaia data release at 1-2 years into the mission.} 
\label{fig:freq_hip_365}
\end{figure}

Nearly every study of classification of periodic variables has used period (or frequency) as a feature. Often in the training set, the period is correct for a large majority of sources due to the investigators selecting the highest quality light curves of each source class of interest. However, if periods are estimated incorrectly for the unlabeled data, then a classifier constructed on the training data may not capture the period--class relationship as it exists for the unlabeled data.

For example, it has been suggested that light curves from early Gaia data releases be labeled using classifiers trained on Hipparcos light curves \citep{2008AIPC.1082..257E,eyer2010gaia}.  Figure \ref{fig:freq_hip_365}a shows a density plot of the estimated frequency for three source classes in Hipparcos\footnote{Sources used in \cite{dubath2011random}} using light curves from the entire 3.5-year survey. The median number of flux measurements per light curve is 91. 
However, one year into \hip{} the densities of the estimated frequency for these source classes look significantly different (Figure \ref{fig:freq_hip_365}b). The median number of flux measurements per light curve is now 29. Thus, even if we assume that Gaia and \hip{} have similar survey characteristics, a classifier built on the 3.5-year baseline \hip{} training set will not accurately capture the frequency--class relationship as it exists in 1-year Gaia data.  This is due to incorrect estimates of frequency for the 1-year length light curves. Since it is often the case that many features depend on frequency (e.g. Table 4 of \cite{richards2011machine} and Section 4.5 of \cite{dubath2011random}), systematic differences in estimates of frequency can alter the distributions of many features.

\subsection{Time-Ordered Flux Measurements}
\label{ss:cadence}

\begin{figure}[ht]
\epsscale{0.29999999999999999}
\begin{center}
\begin{tabular}{ccc}
$\begin{array}{ccc}
\multicolumn{1}{l}{\mbox{(a)}} & \multicolumn{1}{l}{\mbox{(b)}} & \multicolumn{1}{l}{\mbox{(c)}} \\ \\ [-.08in]
\plotone{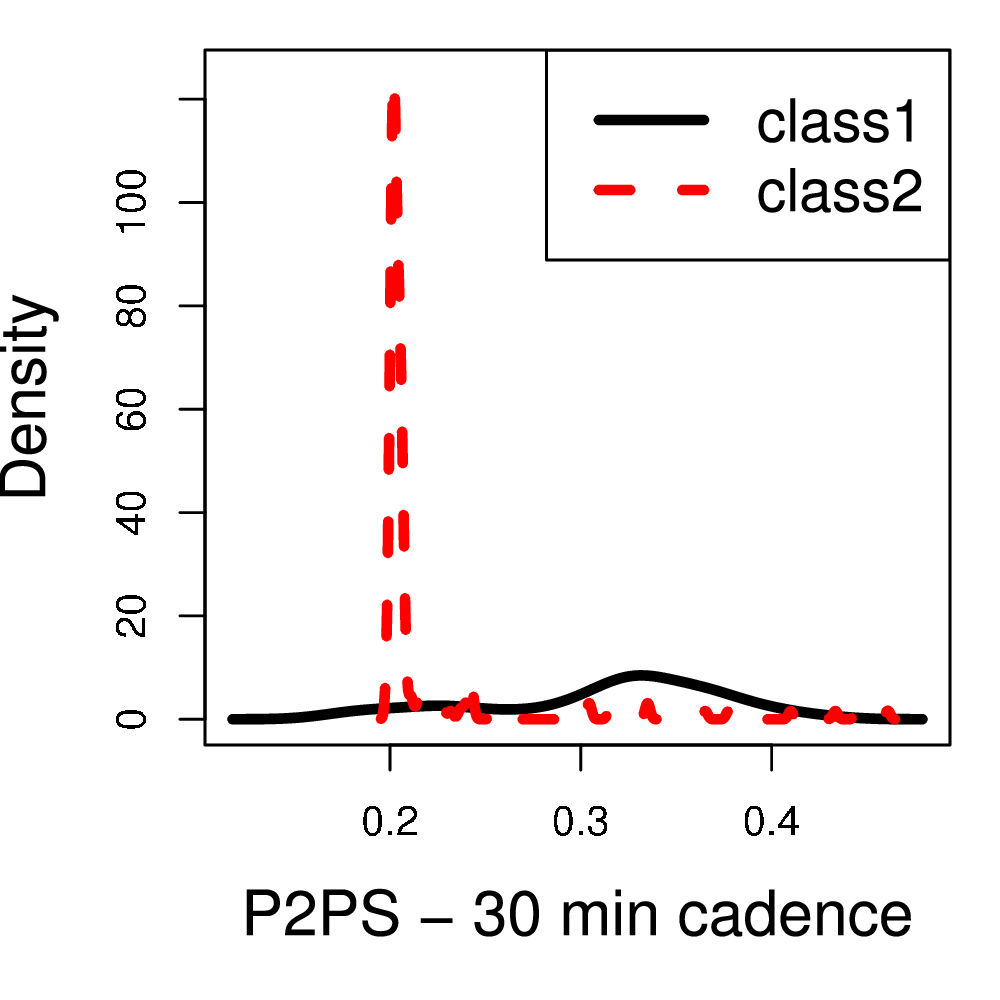} &
\plotone{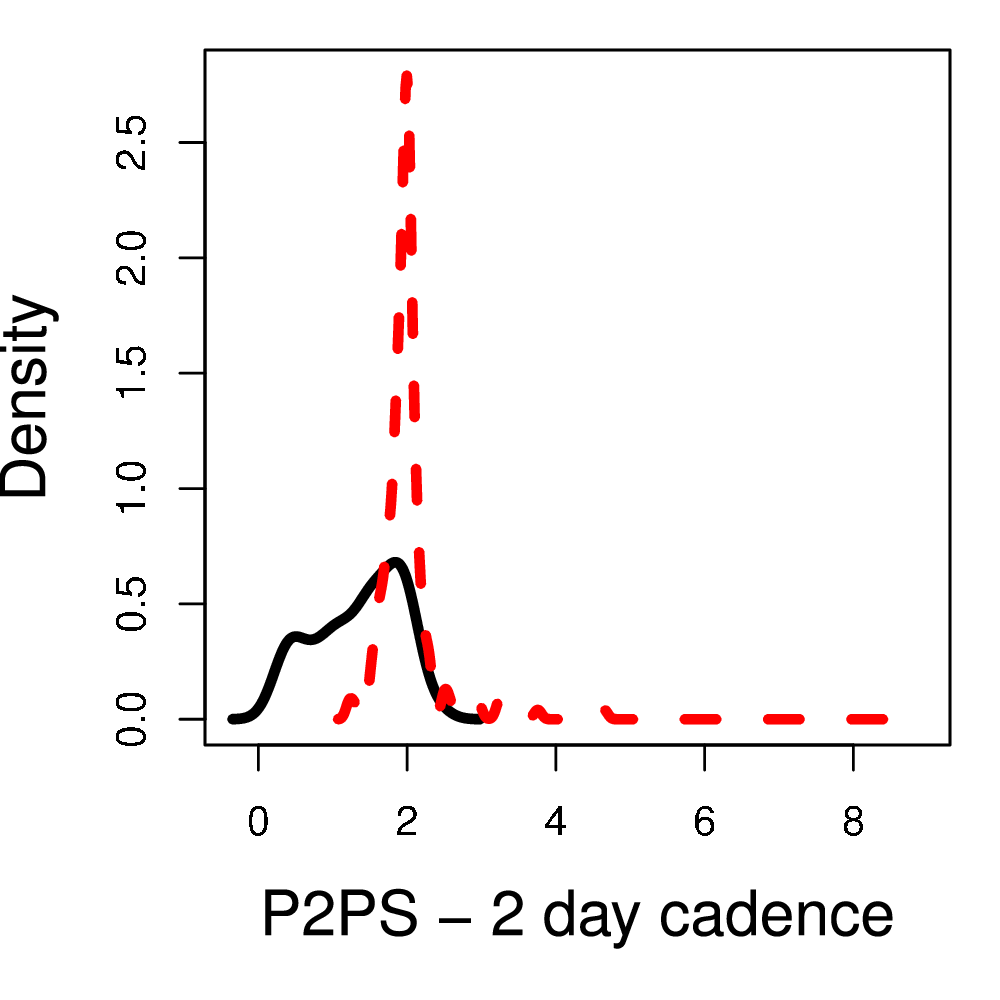} & 
\plotone{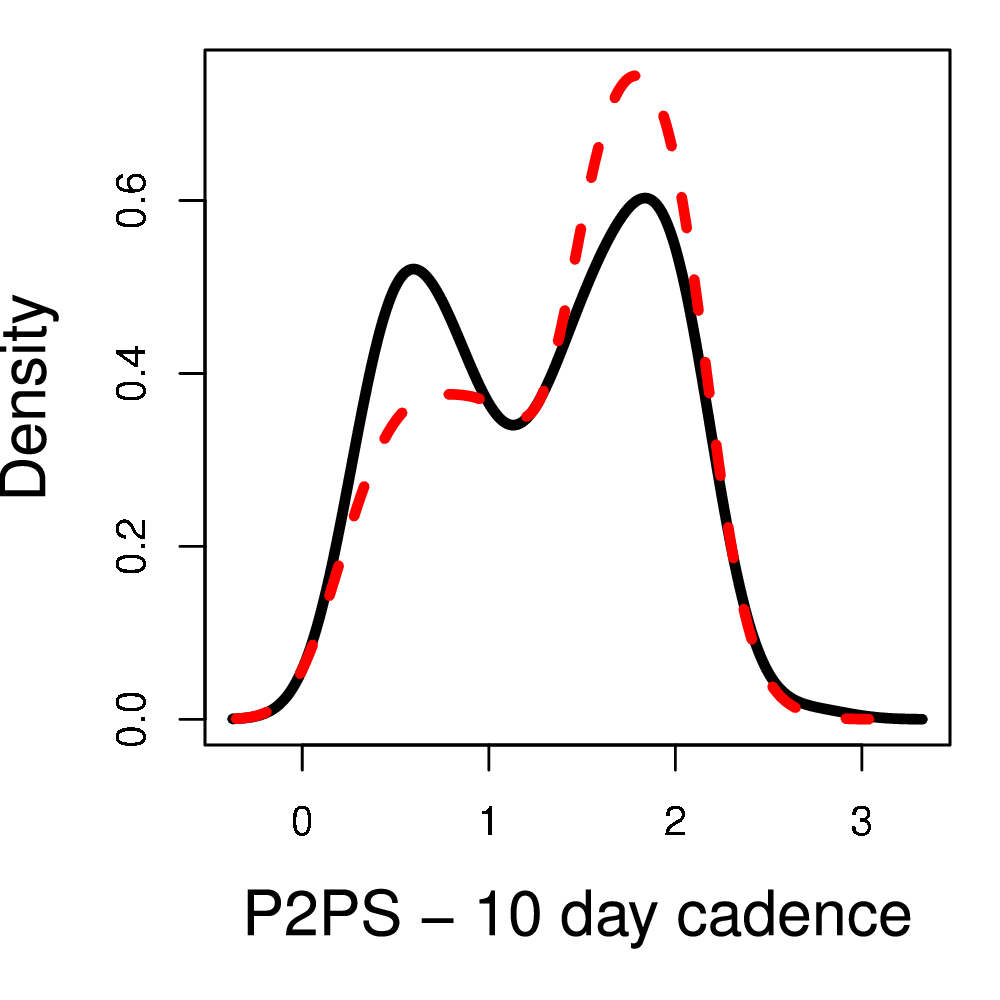}
\end{array}$
\end{tabular}
\end{center}
\caption{Feature distributions can change dramatically with cadence. Plotted are the distributions of the P2PS feature (see equation \eqref{eq:P2PS}) for two simulated classes observed at (a) 30 minute, (b) 2 day, and (c) 10 day cadences. A classifier trained on these light curves at one particular cadence may have poor performance when applied to light curves observed at a different cadence due to this change in feature distribution.\label{fig:dubath_feature20}}
\end{figure}

Several recent studies of classification of periodic variables have used features that depend on the time ordering of flux measurements. For example, \cite{dubath2011random} used \textit{point--to--point scatter} (P2PS), the median of absolute differences between adjacent flux measurements divided by the median absolute difference of flux measurements around the median. Specifically, given some light curve $\textbf{x}$ with time ordered flux measurements $m_0, \ldots m_k$,
\begin{equation}
\label{eq:P2PS}
\text{P2PS}(\mathbf{x}) =  \frac{\mathcal{M}(\{|m_i - m_{i-1}|\}_{i=1}^k)}{\mathcal{M}(\{|m_i - \mathcal{M}(\{m_j\}_{j=0}^k)\}_{i=0}^{k})}
\end{equation}
where $\mathcal{M}$ denotes the median. While potentially useful for classification, the behavior of this feature is heavily dependent on the cadence of time sampling. To see this, consider a two class problem where class 1 is sine waves of amplitude 1 with period drawn uniformly at random between 0.25 days and 0.75 days and class 2 is sine waves of amplitude 1 where period is drawn uniformly at random between 2 days and 8 days. Say we observe 20 flux measurements for each source. Figure \ref{fig:dubath_feature20} shows the density of P2PS for 200 sources of each class with (a) 30 minutes, (b) 2 days, and (c) 10 days between successive flux measurements. At 30 minutes and 2 days the feature is useful for distinguishing classes, but in opposite directions. At 10 days the feature is no longer useful.

The process of how cadence and period produce the P2PS feature density is complex. For class 2 (2 day to 8 day periods) at 30 minute cadence, the flux measurements for each source are often monotonically increasing or decreasing, producing a small numerator relative to denominator in equation \eqref{eq:P2PS}. When the cadence is large relative to the distribution of periods for the source class, the functional shape of the light curve determines the P2PS density. In Figure \ref{fig:dubath_feature20}c where the cadence is longer than any possible period for either class, the two classes have the same density because they have the same functional shape (sine waves).

Note that this extreme sensitivity to cadence is not based on having 20 flux measurements per light curve. Running these simulations with 100 flux measurements per light curve produces densities of roughly the same shape. Rather, this example suggests how useful P2PS may be for distinguishing between classes in a setting where it may be difficult to determine a correct period (20 flux measurements per light curve), and how sensitive it is to systematic differences in cadence between training and unlabeled data.

\subsection{Time-Independent Features}

\begin{figure}[ht]
\epsscale{0.45000000000000001}
\begin{center}
\plotone{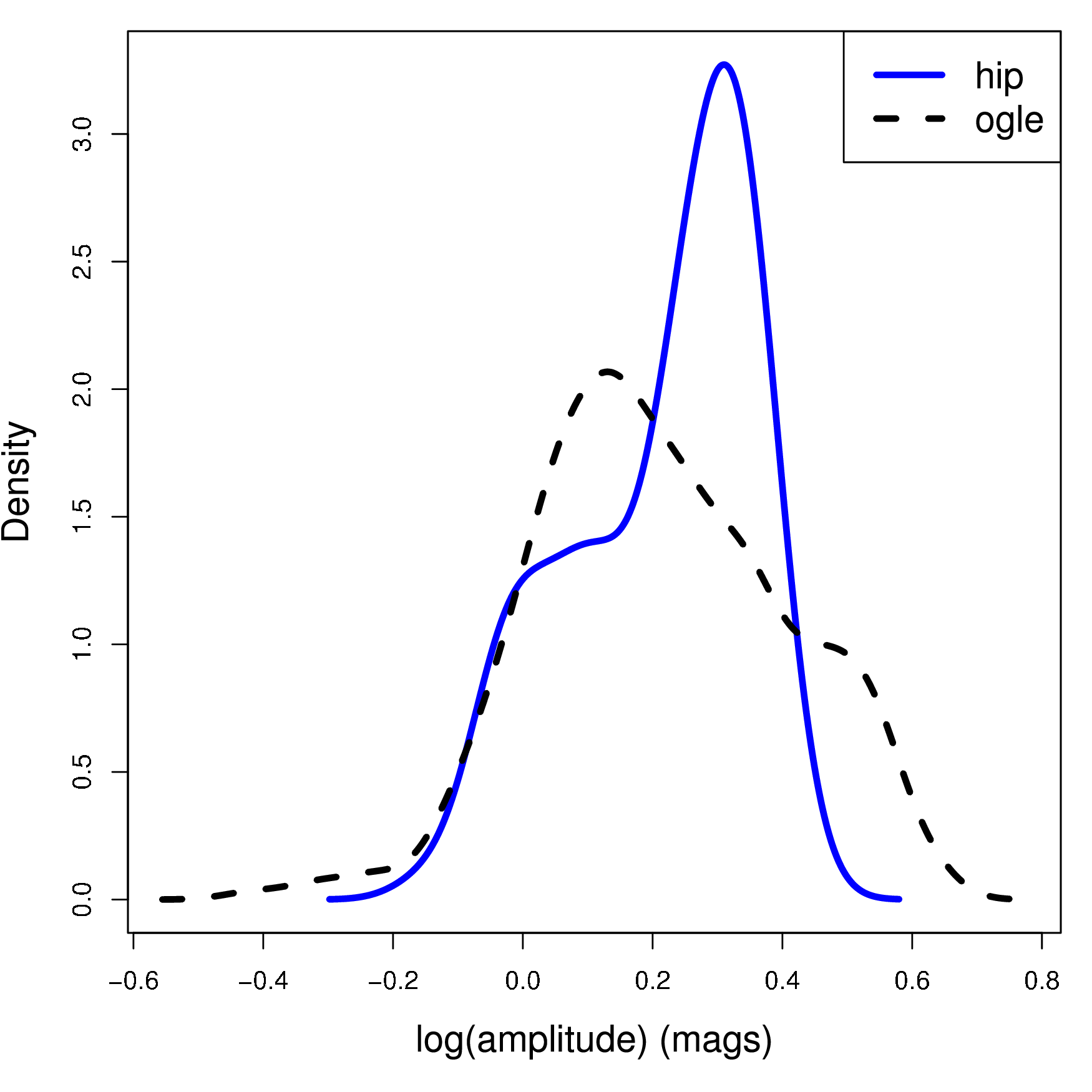}
\end{center}
\caption{Distribution of amplitude for Miras in OGLE and \hip{}. \hip{} Miras do not have very large amplitudes seen in some OGLE sources. The modes of the distributions are different as well.}
\label{fig:amplitude_variation}
\end{figure}

Finally, some of the most useful features for periodic variable classification are simple functions of flux measurements such as estimated amplitude, standard deviation, and skew. Figure \ref{fig:amplitude_variation} shows how estimated amplitude of Miras differs in distribution between the \hip{} and OGLE surveys.\footnote{The \hip{} Miras were used in \citep{debosscher2007automated}. The OGLE sources are V-band data from OGLE III Catalog of Variable Stars: \url{http://ogledb.astrouw.edu.pl/~ogle/CVS/}} In \hip{} there are no Miras with amplitude greater than 3 mag while roughly 12\% of Miras in OGLE have amplitude greater than 3 mag. The mode of the densities is different as well.

There are several possible causes for the difference in shape of these densities. The median difference between last observation time and first observation time for OGLE sources is 1902 days and 1142 days for \hip{}. Since Miras vary in amplitude through each period, it is possible that OGLE is simply observing more periods and picking up on lower troughs and higher peaks than \hip{}. Additionally, many OGLE sources have large mean photometric error (not shown), which may be driving up estimates of amplitude. Also, OGLE and \hip{} sources were observed with different filters, possibly leading to biases in estimated amplitude.

\begin{figure}[ht]
\epsscale{0.45000000000000001}
\begin{center}
\plotone{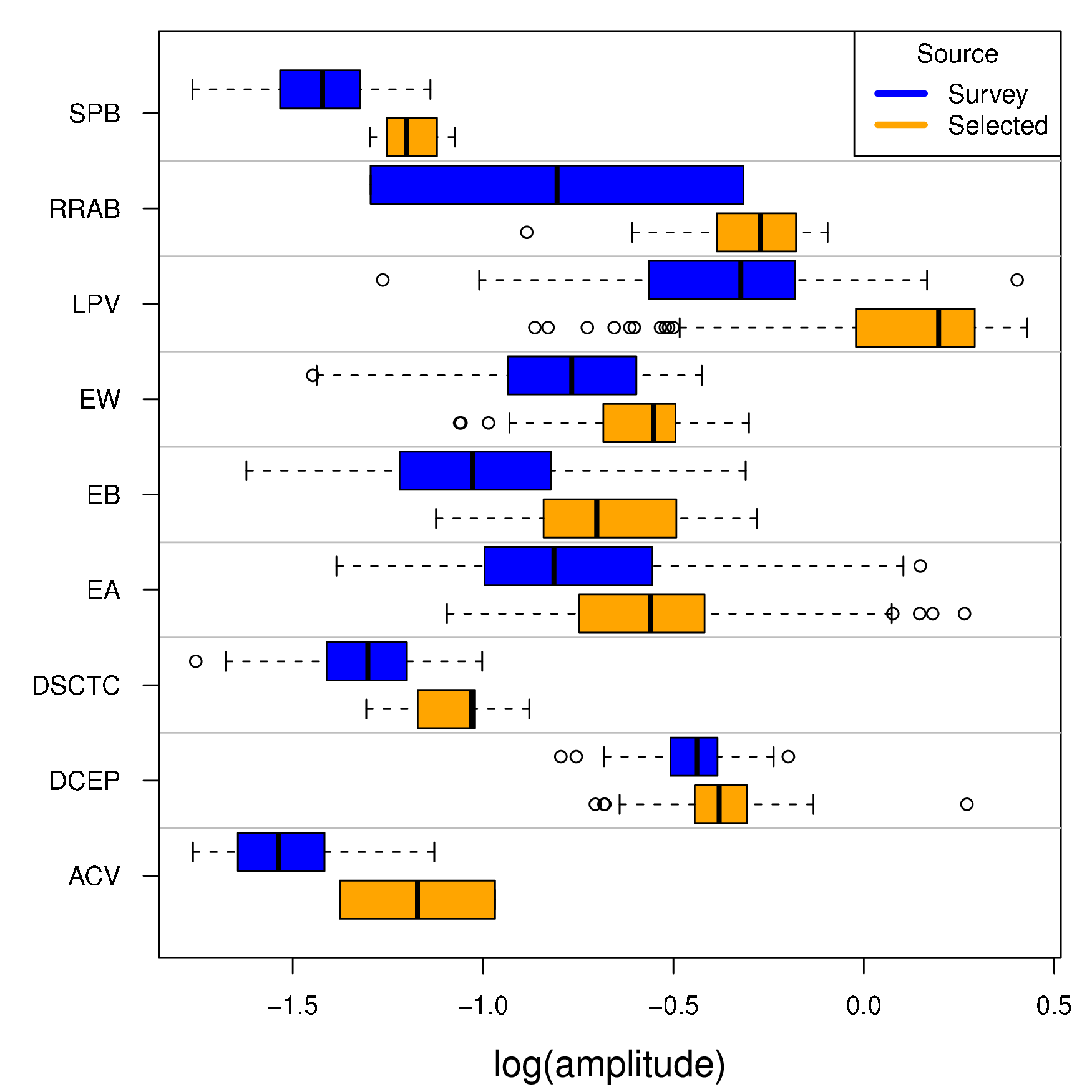}
\end{center}
\caption{Distribution of amplitude for Survey and Selected sources in \hip{}. The Selected sources have systematically larger amplitudes than the Survey sources.}
\label{fig:amplitude_survey_selected}
\end{figure}

It is also worth noting that the \hip{} catalog light curves are themselves a composite of \textit{Selected} sources chosen for their scientific interest before the mission and a set of \textit{Survey} sources which represent a nearly complete sample to well defined magnitude limits (which depend spectral type and galactic latitude). Figure \ref{fig:amplitude_survey_selected} shows boxplots of amplitudes in \hip{} for classes with over 50 sources, blocked into Survey and Selected. The Selected sources appear to have larger amplitudes on average than the Survey sources. A statistical classifier trained on this data will discover class boundaries for this mixture of Selected and Survey sources. However if the unlabeled data resemble the Survey sources, these boundaries may not separate classes well.

\section{Noisification}
\label{sec:noisification}

We have shown how differences in survey systematics can alter feature distributions and deteriorate classifier performance. These survey systematics exist between \textit{and} within surveys. In this section we describe noisification, our solution to addressing training--unlabeled set differences. We use noisification to overcome differences in training--unlabeled feature distributions caused by differences in the number of flux measurements, cadence, and level of photometric error of light curves. Before introducing noisification we discuss a few recent works in the periodic variable classification literature that account for differences in training and unlabeled data and the extent to which they address distribution shifts discussed in Section \ref{sec:feature_distributions}. 

\subsection{Related Work}
\label{sec:lit_review}

Two recent works, \cite{richards2011activelearning} and \cite{debosscher2009automated}, have adapted classifiers to address training--unlabeled data set differences by adding unlabeled data to the training set. \cite{richards2011activelearning} applied an \emph{active learning} methodology to successfully improve classifier performance on ASAS unlabeled data using OGLE and \hip{} training data. \cite{debosscher2009automated} used a method similar to \emph{self-training} \citep{Nigam:2000:AEA:354756.354805} where after applying a classifier trained on \hip{} and OGLE sources to CoRoT data, the most confidently labeled CoRoT sources were added to the training data. From this new training set, they constructed a classifier and used it to classify the remaining CoRoT sources.

Both active learning and self-training are designed to work when the feature densities in training and unlabeled data are different, but the feature--class relationship is the same. More formally, if $p_{tr}(\mathbf{x})$ and $p_{u}(\mathbf{x})$ are the feature densities in training and unlabeled data, then Active Learning and self-training are designed to address the setting where $p_{tr}(\mathbf{x}) \neq p_{u}(\mathbf{x})$, not $p_{tr}(z|\mathbf{x}) \neq p_{u}(z|\mathbf{x})$. However with our problem, differences in number of flux measurements, cadence, and photometric error induce different relationships between class and features. For instance, consider the P2PS cadence example in \S \ref{ss:cadence}, Figure \ref{fig:dubath_feature20}. If the left plot, (a), is the training data P2PS class densities and the center plot, (b), is the unlabeled P2PS class densities, then moving data from (b) to (a) (as is done with Active Learning and self-training) would produce class densities that are a mixture of (a) and (b). Training a classifier on a mixture of (a) and (b) densities is unlikely to produce a classifier that has high accuracy on data with the classes densities in (b).

A method that comes closer to addressing class--feature distribution differences was used in \cite{debosscher2009automated} to overcome aliasing in period estimation. There the authors found that the $13.97^{-1}$ day orbital frequency of the CoRoT mission caused spurious spectral peaks and induced incorrect period estimation for sources. Their solution was to disregard spectral peaks at the orbital frequency. 

Effectively, \cite{debosscher2009automated} asked the question ``What would the value of this light curve's period feature have been if it had been observed at a cadence matching the training data.'' In their case, the answer is fairly staightfoward. However it is much less clear how to correct other features in a similar manner. If the unlabeled sources are observed for 10 days, then it is likely that estimates of amplitude are biased. But by how much? If the source is a Mira, then likely by a lot, but if the source is an RR Lyrae possibly not at all. So in order to correct amplitude estimates we need to know, or have some idea, of the class of the unlabeled source. But this returns to the goal of classification in the first place.

In \cite{long2011classification} this approach was termed \textit{denoisification}. For each unlabeled source the authors estimated a distribution across features representing uncertainty on what the feature values would have been if the source had been observed at a cadence, noise--level, and number of flux measurements in the training data. This distribution was combined with a classifier constructed on training data in order to classify unlabeled sources. While denoisification was superior to not adjusting for training--unlabeled distribution differences, the method did not achieve as large performance increases as \textit{noisification}.

Noisification overcomes training--unlabeled set differences by altering the training set so that the number of flux measurements, cadence, and photometric error match that of the unlabeled data. A classifier can then use this ``noisified'' training data to determine class boundaries as they exist for the unlabeled data. Noisification was introduced in \cite{starr2010map}. \cite{long2011classification} described a specific version of noisification appropriate for when training and unlabeled data have different numbers of flux measurements but are otherwise identical. Here we describe a far more general version of noisification which can be used across surveys when unlabeled sources have a systematically different number of flux measurements, cadence, and photometric error than the training data. Code written in Python and R is available for implementing noisification of light curves.\footnote{Code available here: \url{http://stat.berkeley.edu/~jlong/noisification}}

\subsection{Implementation of Noisification}

Given a set of training light curves, we first estimate a period for each.\footnote{Noisification assumes we have training sources that are of high enough quality that we can estimate periods accurately.} Next, we smooth the period folded light curves, turning each set of flux measurements into a continuous periodic function. Select a light curve $x$ from the training set, and then at random choose a light curve, $l$ from the unlabeled set. Let $g$ be the smooth periodic function associated with $x$. Let $l_{i,1},l_{i,2},$ and $l_{i,3}$ represent the time, flux and photometric error for epoch $i$ of light curve $l$. Say there are $m$ flux measurements for light curve $l$. We now extract flux measurements from the periodic function $g$ matching the cadence and photometric error present in $l$. Specifically, if we let $x_{i,1}, x_{i,2},$ and $x_{i,3}$ be the time, flux, and photometric error of light curve $x$ noisified to light curve $l$, then we have,
\begin{align}
&x_{i,1} = l_{i,1} \\
& x_{i,2} = g(l_{i,1} + \alpha) + \epsilon_i \nonumber \\
&x_{i,3} = l_{i,3} \nonumber
\end{align}
for $i \in \{1, \ldots, m\}$ where
\begin{align*}
&\epsilon_i \sim \text{N}(0,l_{i,3}^2)\\
&\alpha \sim \text{Uniform}[0,p]
\end{align*}
$\alpha$ is a phase offset drawn uniformly at random between 0 and the period of g, $p$. This represents that fact that we are equally likely to start observing a source at any point in its phase. $\epsilon_i$ is the the photometric error added to each flux measurement.

\begin{figure*}[ht]
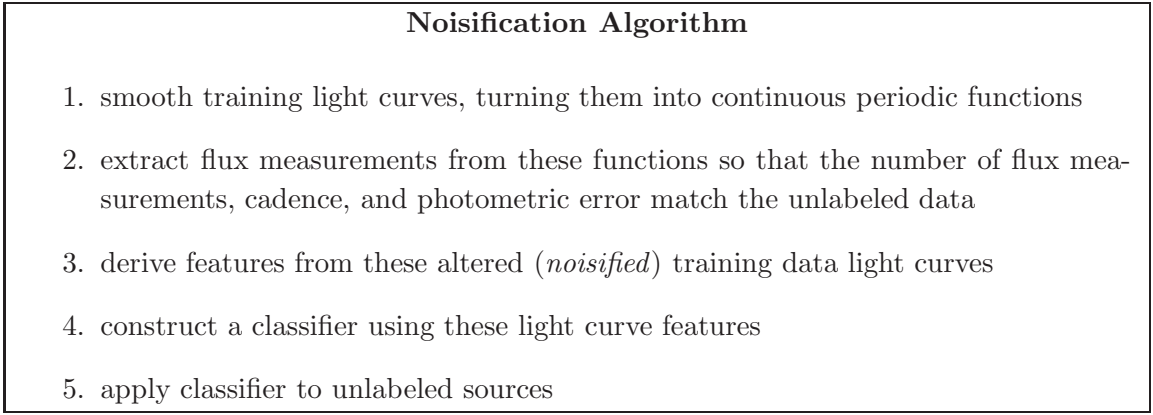

\begin{center}
{\small
\framebox[6in]{
\begin{minipage}[t]{5.8in}
\begin{center} \textbf{Noisification Algorithm} \end{center}
\begin{enumerate}
\item smooth training light curves, turning them into continuous periodic functions
\item extract flux measurements from these functions so that the number of flux measurements, cadence, and photometric error match the unlabeled data
\item derive features from these altered (\textit{noisified}) training data light curves
\item construct a classifier using these light curve features
\item apply classifier to unlabeled sources
\end{enumerate}
\end{minipage}
}
}
\end{center}
\caption{Description of the light curve noisification algorithm.\label{fig:noisification_algorithm}}
\label{lb}
\end{figure*}

The cadence and level of photometric error in this new, noisified version of light curve $x$ now match that of the unlabeled data. Repeat this process for every training light curve. Then derive features for the noisified training data, train a classifier on these observations, and classify the unlabeled light curves using this classifier. We call this process noisification because if our training data consists only of well-sampled light curves and our unlabeled data consists mainly of poorly sampled light curves, then the technique effectively adds noise to features in the training data to more closely match the characteristics of the unlabeled features. See Figure \ref{fig:noisification_algorithm} for a concise description of the algorithm.

\subsection{Remarks on Noisification}
There are a few important points to note about this procedure. First, if the training and unlabeled data have the same cadence and photometric error, then smoothing the training light curves is not necessary. This would be the case, for example, if we had a set of training light curves of known class with many flux measurements ($\sim$ 100) from one survey and we wanted to classify an unlabeled set of poorly sampled light curves ($\sim 30$ flux measurements) of similar cadence and photometric error level from the same survey as the training data. Then we could simply take the training light curves, truncate them at 30 flux measurements, train a classifier on the truncated curves, and apply this classifier to the unlabeled light curves. This setting has the added benefit that no error will be introduced by smoothing the light curves. In this case the training sources do not need to be periodic.

Secondly, the procedure as described is most appropriate if all of the unlabeled data have similar numbers of flux measurements, cadence and photometric error. If this is not the case, then we can repeat the procedure several times using different subsets of the unlabeled data which share similar properties. For example, if unlabeled light curves have either around 20 or around 70 flux measurements, then we could break the unlabeled data into two sets and classify each set using a separate run of the noisification procedure. The more subsets of the unlabeled data one uses, the closer the noisified training data gets to the unlabeled data. The tradeoff is computational burden. With $n$ training light curves and $m$ unlabeled light curves, noisifying to precisely match the properties of each unlabeled light curve requires deriving features for $nm$ light curves. In Section \ref{sec:data_experiments} we explore how much one can gain from dividing the unlabeled data into subsets.

With noisification, the unlabeled light curve, $l$, at which to noisify training light curve $x$, $\alpha$ and $\epsilon$ are all random. Thus, repeating the noisification process several times and obtaining several classifiers offers potential for improvement in classifier performance over running the process once. We study this in Section \ref{sec:data_experiments}. While building several classifiers may be a good idea, it is important not to train a classifier using several noisified versions of the same light curve as the training data would no longer be independent. This can cause classifiers to overfit the data, hurting classifier performance.

Note that noisification is classifier independent. We use Random Forests in this work, but noisification can be used in conjunction with essentially any statistical classification method. Here we use Super Smoother for transforming training light curves into continuous periodic functions \citep{friedman1984variable}\footnote{Fortran code here: \url{http://www-stat.stanford.edu/~jhf/ftp/progs/supsmu.f}. We used automatic span selection (span$=0.0$) and a high frequency penalty of $\alpha=1.0$. These choices were based on visual inspection of smoothing fits to light curves.}. The method used for inferring continuous training curves is separate from the the rest of the noisification process. Splines and Nadaraya-Watson methods are other possibilities. Splines are described in 5.4 of \cite{hastie2009elements}. See \cite{hall2008nonparametric} for using Nadaraya-Watson with periodic variables.

Finally we stress that this implementation of noisification is limited to addressing differences between training and unlabeled sets caused by number of flux measurements, cadence, and photometric error. We do not correct for differences in feature distributions due to observing regions, detection limits, or filters.

\section{Experiments}
\label{sec:data_experiments}
\subsection{Noisification within a Survey}
\label{sec:no_smoothing}

{\small
\begin{table}
\begin{center}
\begin{tabular}{ccccc}
\hline
\textbf{Survey} & \textbf{Source Classes}$^a$ &  \textbf{F / LC}$^b$ & \textbf{\# Train} & \textbf{\# Unlabeled}\\
\hline
Simulated & RR Lyrae, Cepheid, $\beta$ Persei,  & 200-200 & 500 & 500\\
& $\beta$ Lyrae, Mira & & &\\
OGLE$^c$ & RR Lyrae DM, MM Cepheid,  & 261-474 & 358 & 165\\
&   $\beta$ Persei, $\beta$ Lyrae, WU Majoris & & &
\end{tabular}
\caption{Light curves used in Sections \ref{sec:no_smoothing} and \ref{sec:smoothing}.\label{tab:data_sets}} 
\end{center}
{\tiny
$^a$ In the case of the simulated data, the light curves were made to resemble these classes. \\
$^b$ F / LC is the first and third quartiles of flux measurements per light curve for training. \\
$^c$ We use every light curves of these classes analyzed in \cite{richards2011machine}.
}
\end{table}
}

To get a sense how noisification performs in a controlled setting, we first test the method using training and unlabeled data from the same survey, but with systematically differing number of flux measurements. This resembles the real--life situation where well sampled light curves of known class are used as training data to classify poorly sampled curves of unknown class from the same survey. The cadence and levels of photometric error are assumed to match in the training and unlabeled data. We are also free from worrying about survey characteristics that noisification does not address. We perform two experiments, one using a simulated light curve data set and one using an OGLE light curve data set.\footnote{Here the OGLE curves are in I-band.} See Table \ref{tab:data_sets} for data set information.

\begin{figure}[ht]
\epsscale{0.45000000000000001}
\begin{center}
$\begin{array}{cc}
\multicolumn{1}{l}{\mbox{(a)}} & \multicolumn{1}{l}{\mbox{(b)}} \\ \\ \\ [-.35in]
\plotone{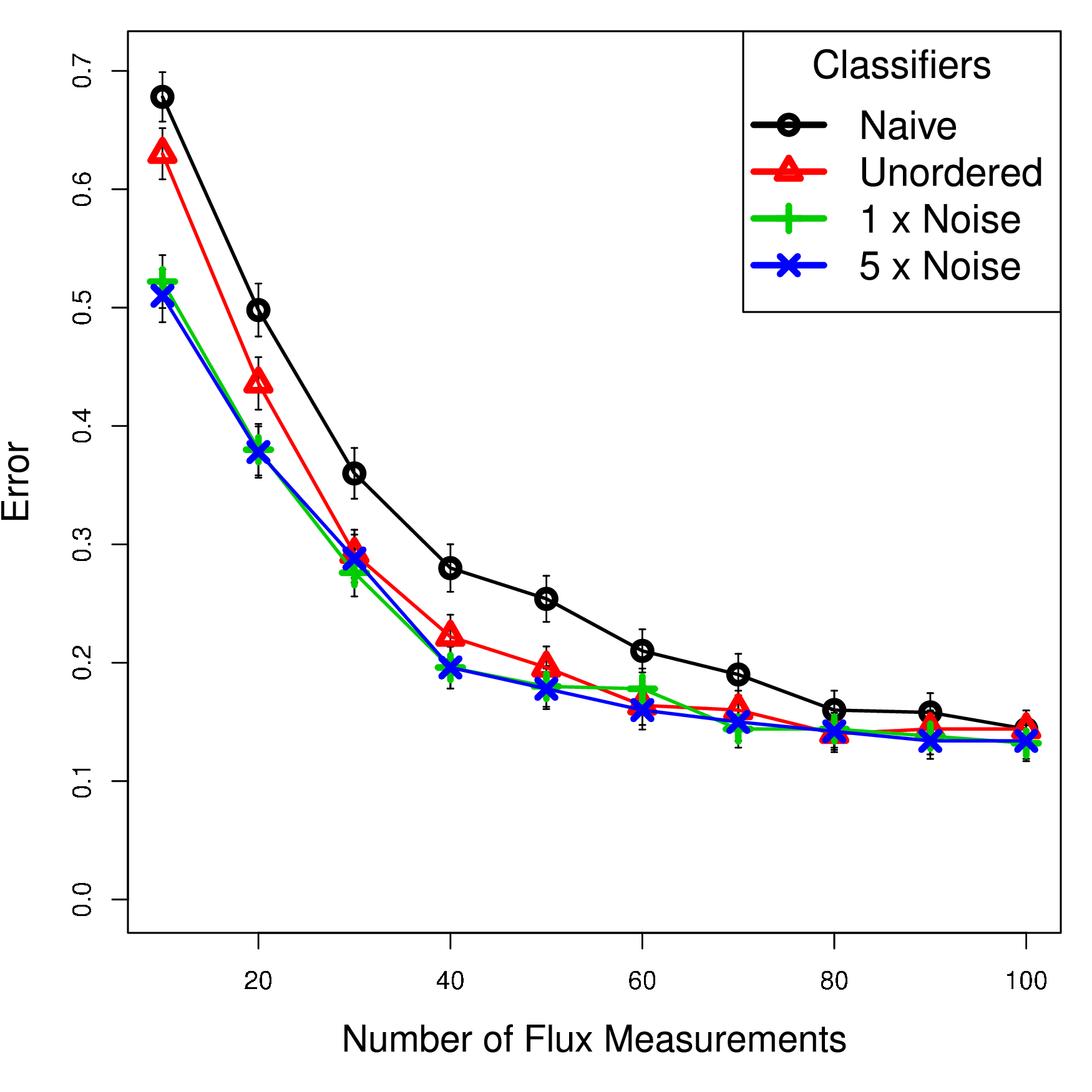} &
\plotone{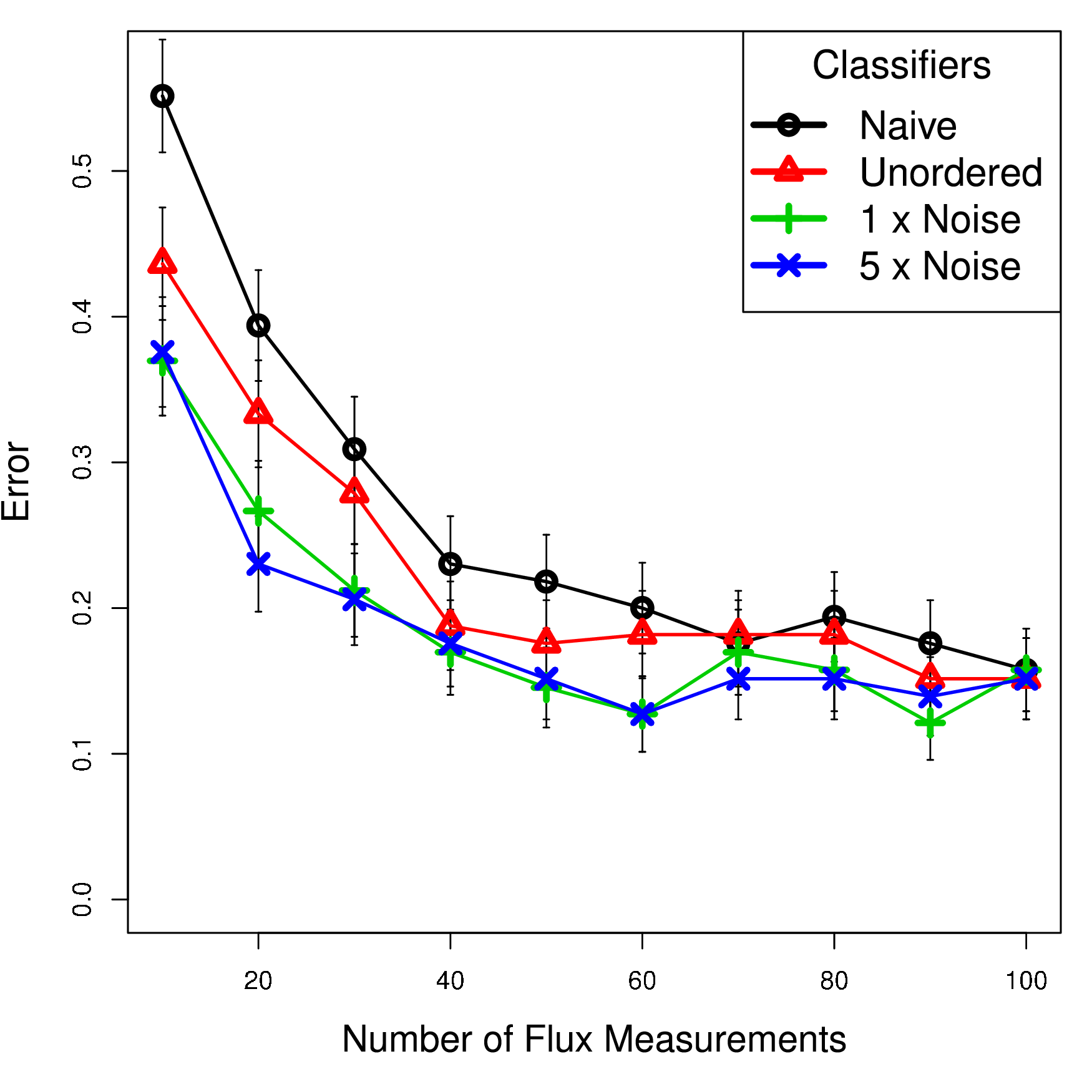}
\end{array}$
\end{center}
\caption{Noisification results for (a) simulated light curves and (b) OGLE light curves. 5x Noisification (blue x) improves over making no adjustments for training--unlabeled data set differences (black circles) at all numbers of flux measurements.}
\label{fig:no_smoothing}
\end{figure}

After splitting each data set into training and unlabeled sets, we downsample the light curves in the unlabeled data set to 10 through 100 flux measurements in multiples of 10. Now the unlabeled data sets resemble the training in every way except for the number of flux measurements per light curve. To each of the ten unlabeled data sets we apply four classifiers and compute classification accuracy on the unlabeled data sets. Figure \ref{fig:no_smoothing} provides error rates for the four classifiers applied to the 10 unlabeled sets from (a) simulated and (b) OGLE. The four classifiers are:
\begin{enumerate}
\item \textbf{naive} (black circles): Random Forest constructed on the unaltered training data
\item \textbf{unordered} (red triangles): noisify every training light curve by matching the number of flux measurements in the training set and unlabeled set, but we choose a random, non-contiguous set of epochs (cadence information is lost)
\item \textbf{1x noisification} (green plus): noisification without smoothing as described in Section \ref{sec:noisification}
\item \textbf{5x noisification} (blue x) ``1x noisification'' repeated five times as discussed in Section \ref{sec:noisification}
\end{enumerate}

The results in Figure \ref{fig:no_smoothing} suggest that noisification can significantly increase classification performance when the unlabeled data is poorly sampled. With OGLE, ``naive'' misclassifies around 32\% of light curves with 30 flux measurements while ``5x noisification'' misclassifies around 21\%. Based on the difference between the ``unordered'' and ``1x / 5x noisification'' procedures, it appears that having a training cadence that matches the cadence of the unlabeled data can improve classification performance. We explore this in more detail later when training and unlabeled data come from surveys with different cadences. The ``5x noisification'' advantage over ``1x noisification'' is fairly modest. Repeatedly noisifying the training data and averaging the resulting classifiers reduces variance and leaves bias unchanged, so we see no way that using ``5x noisification'' instead of ``1x noisification'' could hurt classifier performance. For the remainder of the paper, noisification refers to``5x noisification.''

\begin{figure}[ht]
\epsscale{0.45000000000000001}
\begin{center}
\begin{tabular}{cc}
$\begin{array}{cc}
\multicolumn{1}{l}{\mbox{(a)}} & \multicolumn{1}{l}{\mbox{(b)}} \\ \\ \\ [-.35in]
\plotone{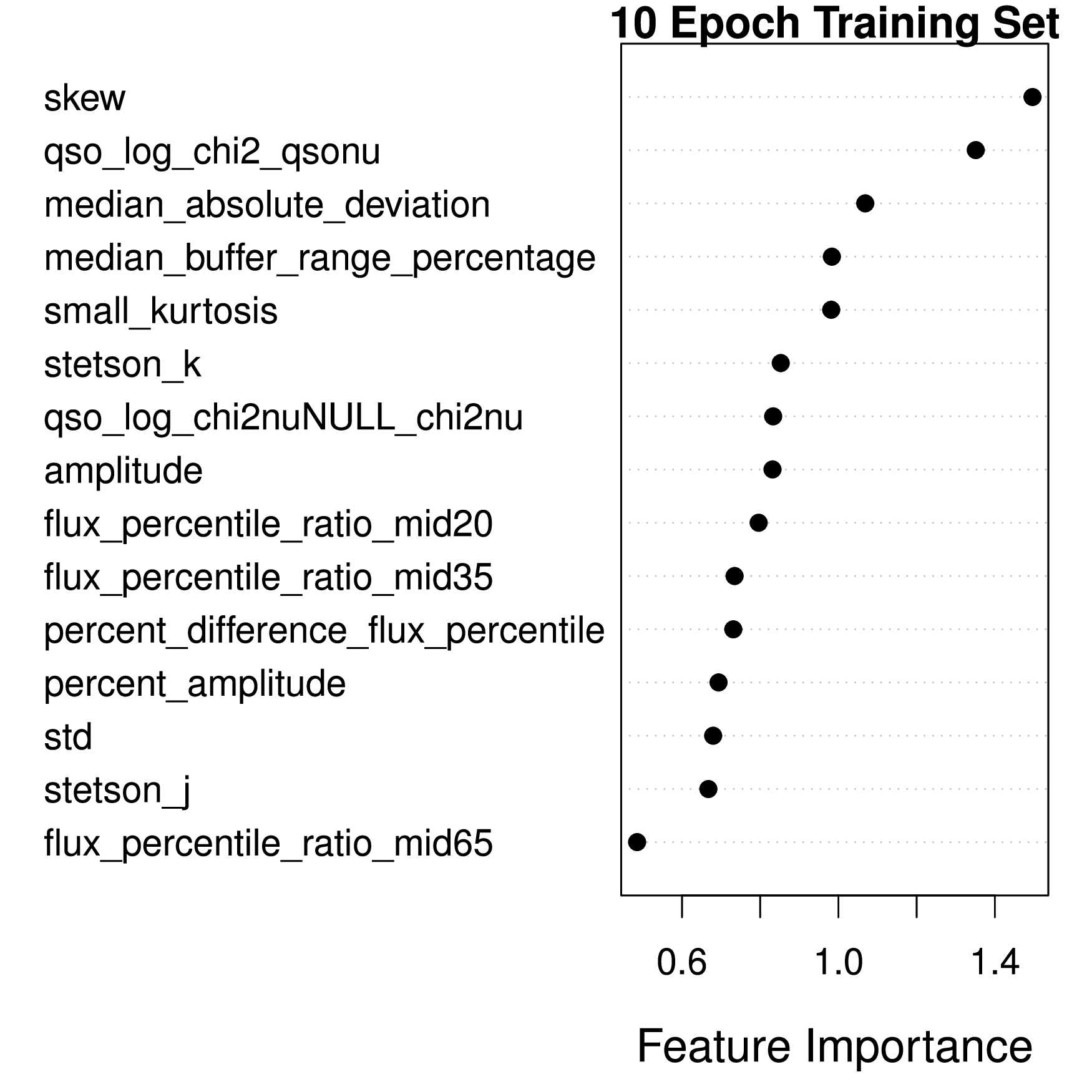} &
\plotone{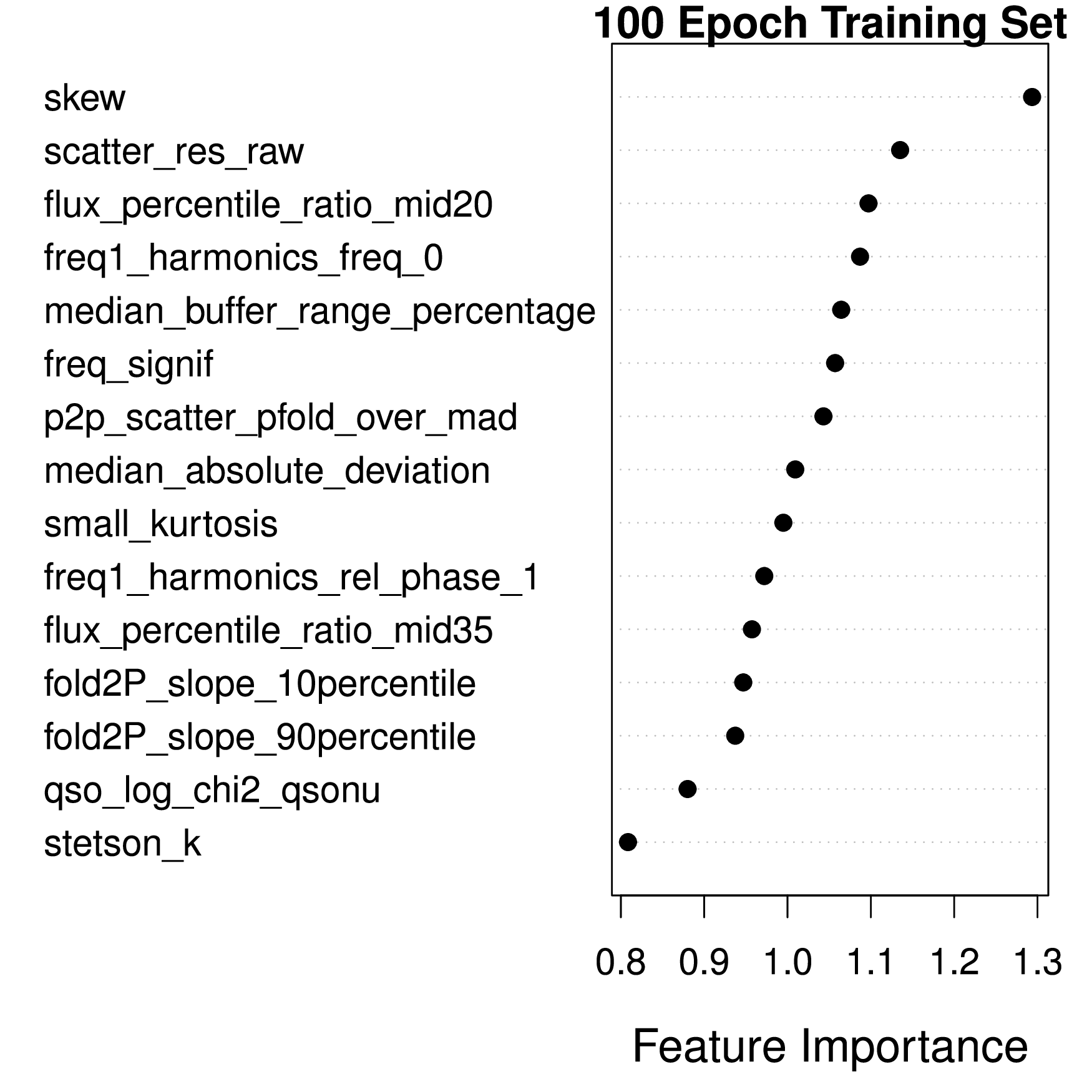}
\end{array}$
\end{tabular}
\end{center}
\caption{Variable importances for the OGLE ``1x noisified'' classifier on (a) 10 flux measurement and (b) 100 flux measurement training sets. When the training data has few flux measurements non-periodic features are most important because periods cannot be estimated correctly. See Section 4.2 of \cite{dubath2011random} for an explanation of feature importance.}
\label{fig:var_imp}
\end{figure}

To investigate how noisified classifiers differ, we plot feature importances for the ``1x noisification'' classifier for 10 and 100 flux measurements for the OGLE data (see Figure \ref{fig:var_imp}). Random Forest feature importance measures were introduced by \cite{breiman2001random} and have been used in recent studies of periodic variables to gain an understanding of which features Random Forests considers most highly when assigning a class to a light curve. See \cite{dubath2011random} Section 4.1 for a complete description of feature importance. Figure \ref{fig:var_imp} shows that skew is very important for both classifiers. Notice that the 100 flux measurement classifier ranks several period based features as being important -- \textit{scatter\_res\_raw}, \textit{freq\_signif}, and \textit{freq1\_harmonics\_freq\_0} -- while the 10 flux measurement classifier does not. The interpretation is clear: when classifying light curves with 10 flux measurements, features that require a correct period will not be very useful. The process of noisifying light curves causes the classifier to recognize this and make use of class information present in other features.

\begin{figure}[ht]
\epsscale{0.45000000000000001}
\begin{center}
\begin{tabular}{cc}
$\begin{array}{cc}
\multicolumn{1}{l}{\mbox{(a)}} & \multicolumn{1}{l}{\mbox{(b)}} \\ \\ \\ [-.35in]
\plotone{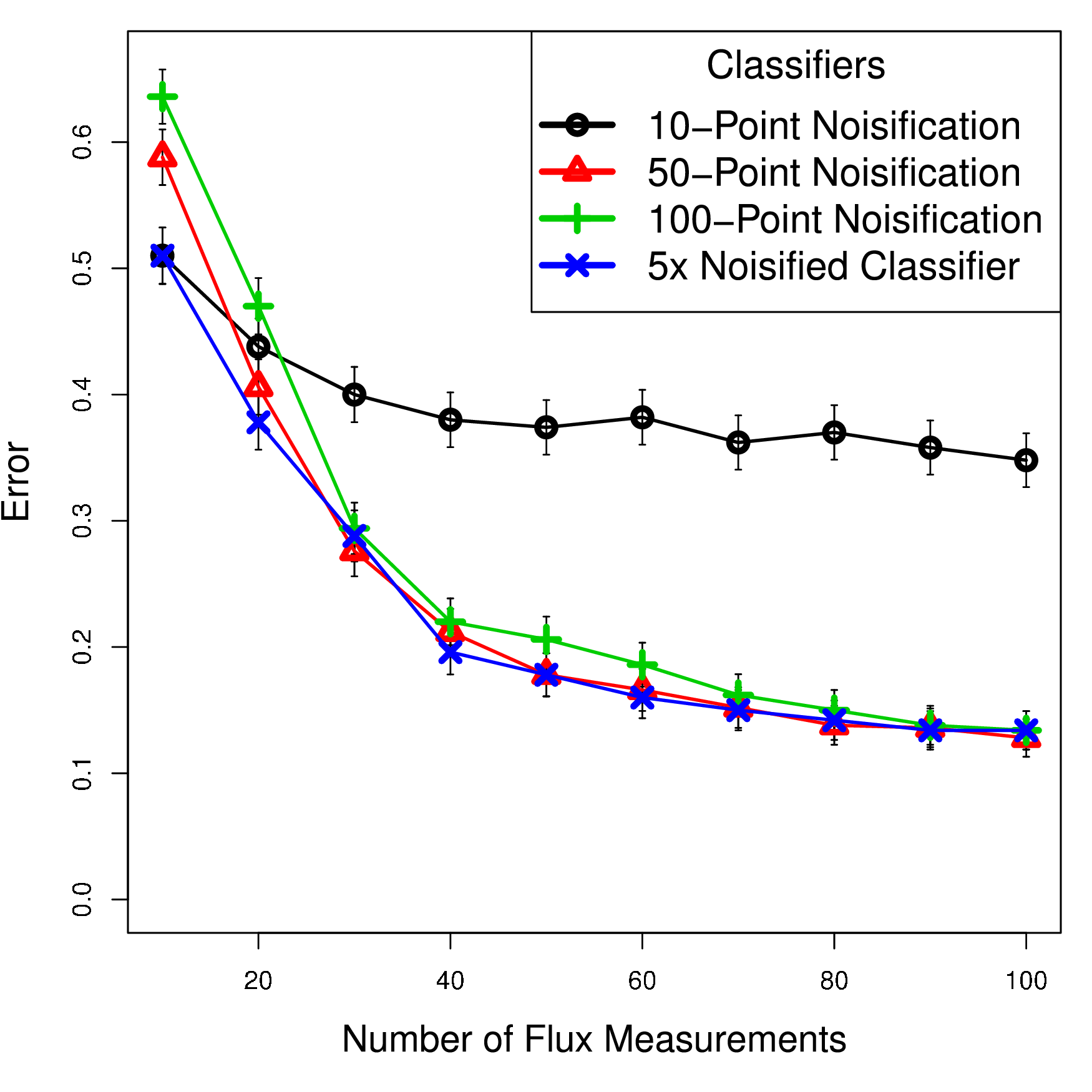} &
\plotone{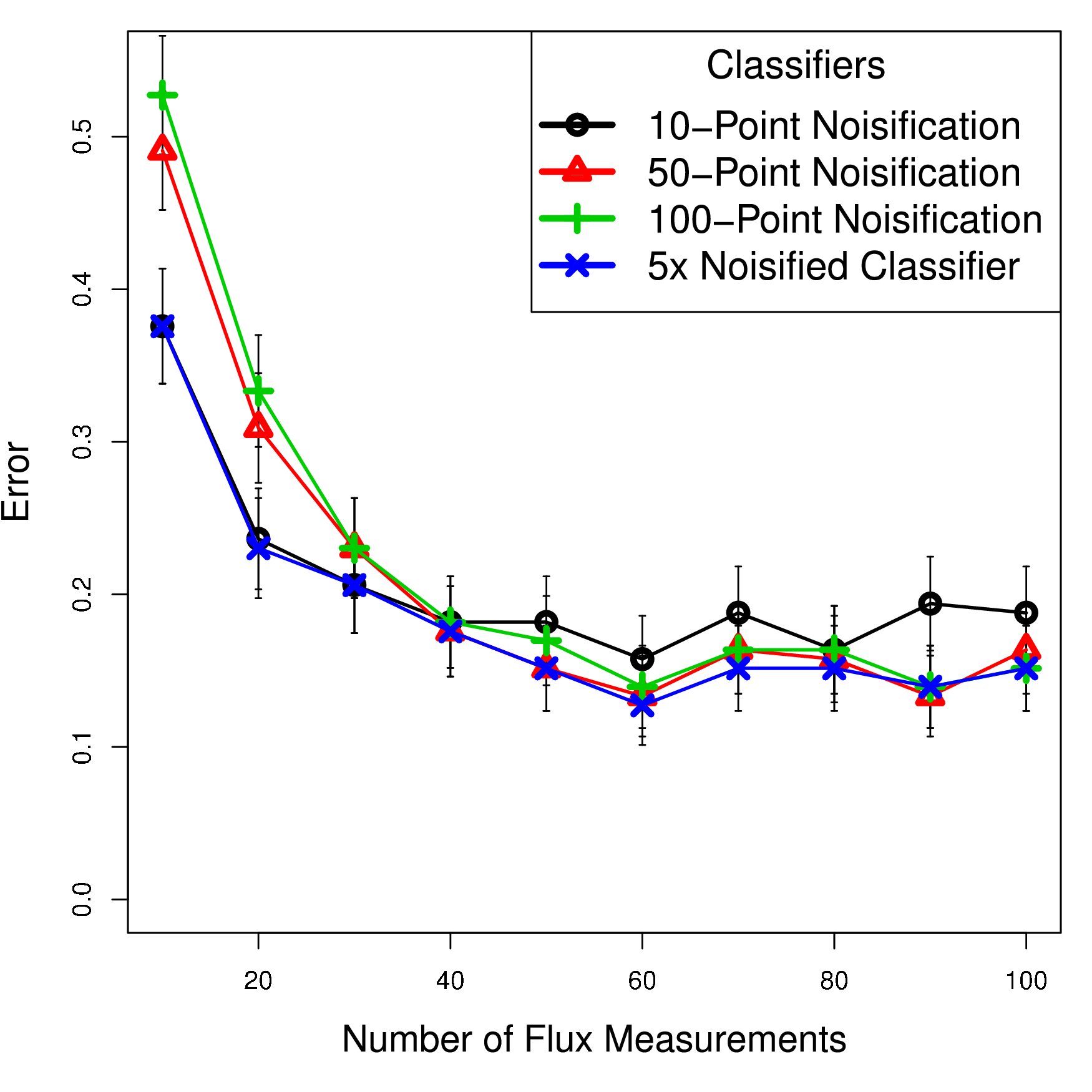}
\end{array}$
\end{tabular}
\end{center}
\caption{The 10-point, 50-point, and 100-point noisified classifiers applied to all of the (a) simulated and (b) OGLE unlabeled sets. The 50-point and 100-point noisified classifiers perform well on all the unlabeled data sets with more than 30 flux measurements for both simulated and OGLE.\label{fig:robustError}}
\label{fig:robust_error}
\end{figure}

In these two examples, light curves in the unlabeled data set always had one of 10 possible number of flux measurements ($10, 20, \ldots 100$). The noisified light curves had exactly the same number of flux measurements as the unlabeled data. In practice, we will need to classify light curves with any number of flux measurements. It may be computationally challenging to construct noisified classifiers for every possible number of flux measurements. To test how sensitive error rates are to how light curves are noisified, we took the noisified classifiers for 10, 50, and 100 flux measurements and applied them across all 10 of the unlabeled data sets. Figure \ref{fig:robustError} shows the results for the (a) simulated and (b) OGLE data. We plot the error rates of these three classifiers along with the error rate of the classifier noisified to the number of flux measurements actually in the unlabeled data set (the ``5x noisified'' classifiers from Figure \ref{fig:no_smoothing}). The results show that for these examples the error rates are fairly insensitive to exactly how many flux measurements we use in the noisified classifier. For the OGLE data, the classifier noisified to 10 flux measurements performs well until unlabeled light curves have around 70 flux measurements. Additionally the 50--flux and 100--flux noisified classifiers perform well for unlabeled data sets with between 30 and 100 flux measurements.

\subsection{Noisification with Smoothing}
\label{sec:smoothing}

We now address the challenge of training a classifier on a survey with one cadence to classify light curves of a different cadence. In order to ensure that all differences between training and unlabeled data are due to issues addressed by noisification (number of flux measurements, cadence, photometric error) we use the simulated light curve prototypes from Section \ref{sec:no_smoothing} for both training and unlabeled data sets. We sample the light curves at actual \hip{} and OGLE light curve cadences used in previous studies \citep{richards2011machine,debosscher2007automated}.

Systematic differences exist between the OGLE and \hip{} survey cadences. OGLE is a ground based survey with flux measurements taken at multiples of one day plus or minus a few hours. The sampling for these curves is quite regular with few large gaps. In contrast, \hip{} light curves tend to be sampled in bursts, with several measurements over the course of less than a day followed by long gaps.

In practice, one data set (say, \hip{}) would be used to train a classifier in order to classify sources in the other (say, OGLE). However since these light curves are simulated, and we have labels for both sets, we create training and unlabeled data sets at \hip{} and OGLE cadences so we can study the challenge of constructing a classifier on \hip{} for use on OGLE sources and vice versa. We begin by generating 1000 simulated light curves using the class templates from Section \ref{sec:no_smoothing}. For 500 of these curves we randomly select an OGLE cadence and sample flux measurements and photometric errors from this cadence. We then take these 500 curves and downsample them to have $10, \ldots, 100$ flux measurements in multiples of 10. The original 500 curves cadenced to OGLE is the OGLE training set, and the downsampled curves are the 10 OGLE unlabeled data sets. We repeat this process for the other 500 simulated curves at \hip{} cadences.

\begin{figure}[ht]
\epsscale{0.45000000000000001}
\begin{center}
$\begin{array}{cc}
\multicolumn{1}{l}{\mbox{(a)}} & \multicolumn{1}{l}{\mbox{(b)}} \\ \\ \\ [-.35in]
\plotone{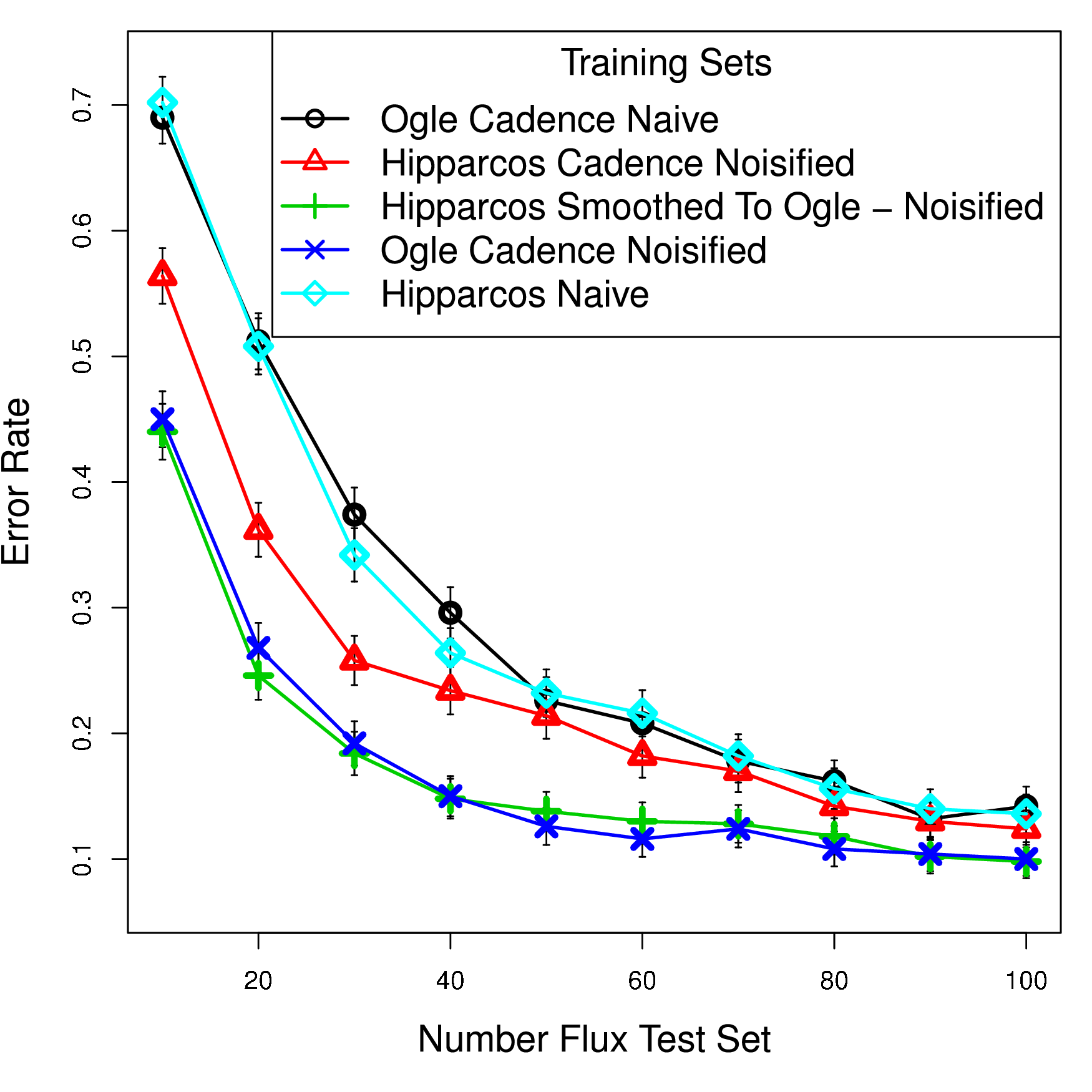} &
\plotone{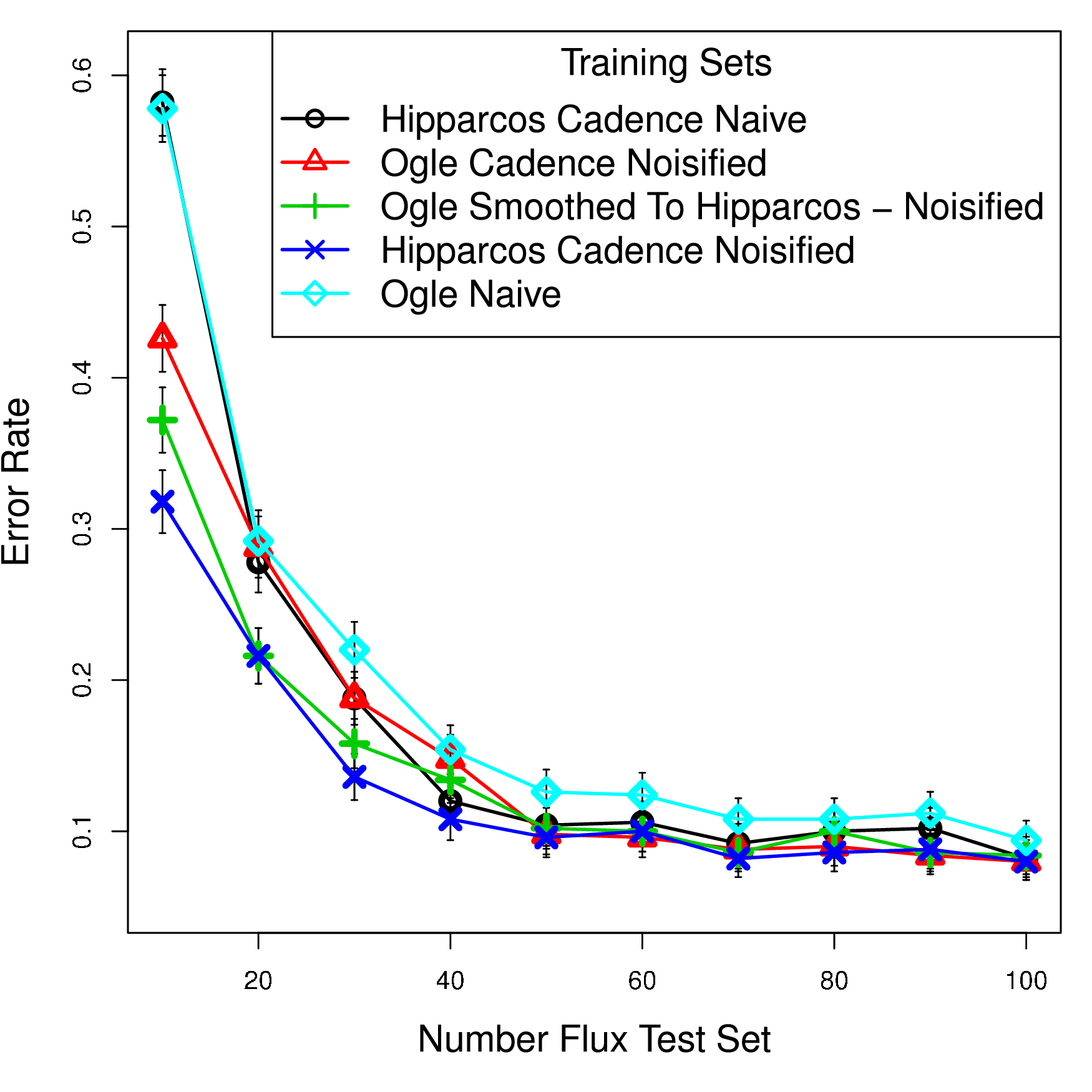}
\end{array}$
\end{center}
\caption{Simulated light curves where the unlabeled data is observed at a (a) OGLE or (b) \hip{} cadence. By smoothing the training set and extracting flux measurements to match that of the unlabeled data (green plus), we improve performance over only matching number of flux measurements (red triangle).}
\label{fig:simulated_smoothing}
\end{figure}

In order to test the efficacy and necessity of various aspects of the noisification process, we apply several classifiers to each of the unlabeled data sets. Figure \ref{fig:simulated_smoothing} shows the accuracy of these methods treating (a) OGLE and (b) \hip{}  as the unlabeled data. For the left plot with OGLE unlabeled light curves the classifiers are trained on:
\begin{enumerate}
\item \textbf{ogle cadence naive} (black circle): unaltered OGLE light curves
\item \textbf{hipparcos cadence noisified} (red triangle): \hip{} light curves truncated to match length of unlabeled set, but not smoothed (cadence is different between training and unlabeled)
\item \textbf{hipparcos smoothed to ogle -- noisified} (green plus): \hip{} light curves after they have been smoothed, cadenced at OGLE, and truncated to match length of unlabeled curves
\item \textbf{ogle cadence noisified} (dark blue x): noisified OGLE light curves (cadence already matches unlabeled set so smoothing unnecessary)
\item \textbf{hipparcos naive} (light blue diamonds): unaltered \hip{} light curves
\end{enumerate}

Not addressing cadence, flux measurement, and photometric error mismatches by training on full length \hip{} light curves leads to poor performance (light blue diamond). Noisifying these \hip{} sources by truncation improves performance (red diamonds). However we gain significantly by correcting for cadence differences by smoothing (green plus). It is encouraging to see that by smoothing the \hip{} training set and noisifying we can do as well as if we had started with OGLE cadence curves (dark blue x and green plus). 

The right plot of Figure \ref{fig:simulated_smoothing} displays the same information with \hip{} as the unlabeled cadence. Note that the line markings have been changed to preserve relationship of training set to unlabeled set. The overall picture is similar to the OGLE data, except that convergence of error rates happens much more quickly. At 60 flux measurements there is little difference among any of the classifiers.

\begin{figure}[ht]
\epsscale{0.45000000000000001}
\begin{center}
$\begin{array}{cc}
\multicolumn{1}{l}{\mbox{(a)}} & \multicolumn{1}{l}{\mbox{(b)}} \\ \\ \\ [-.35in]
\plotone{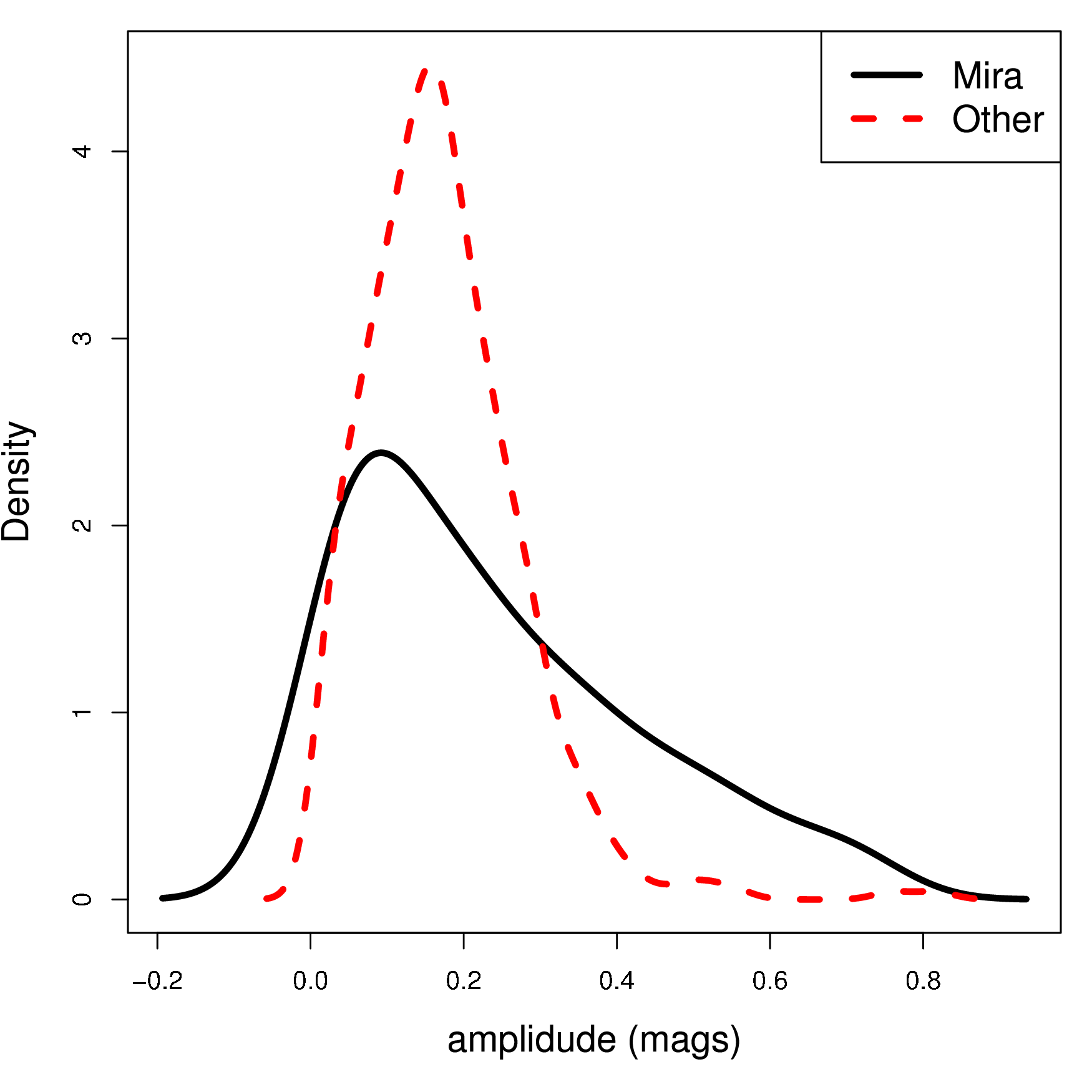} & 
\plotone{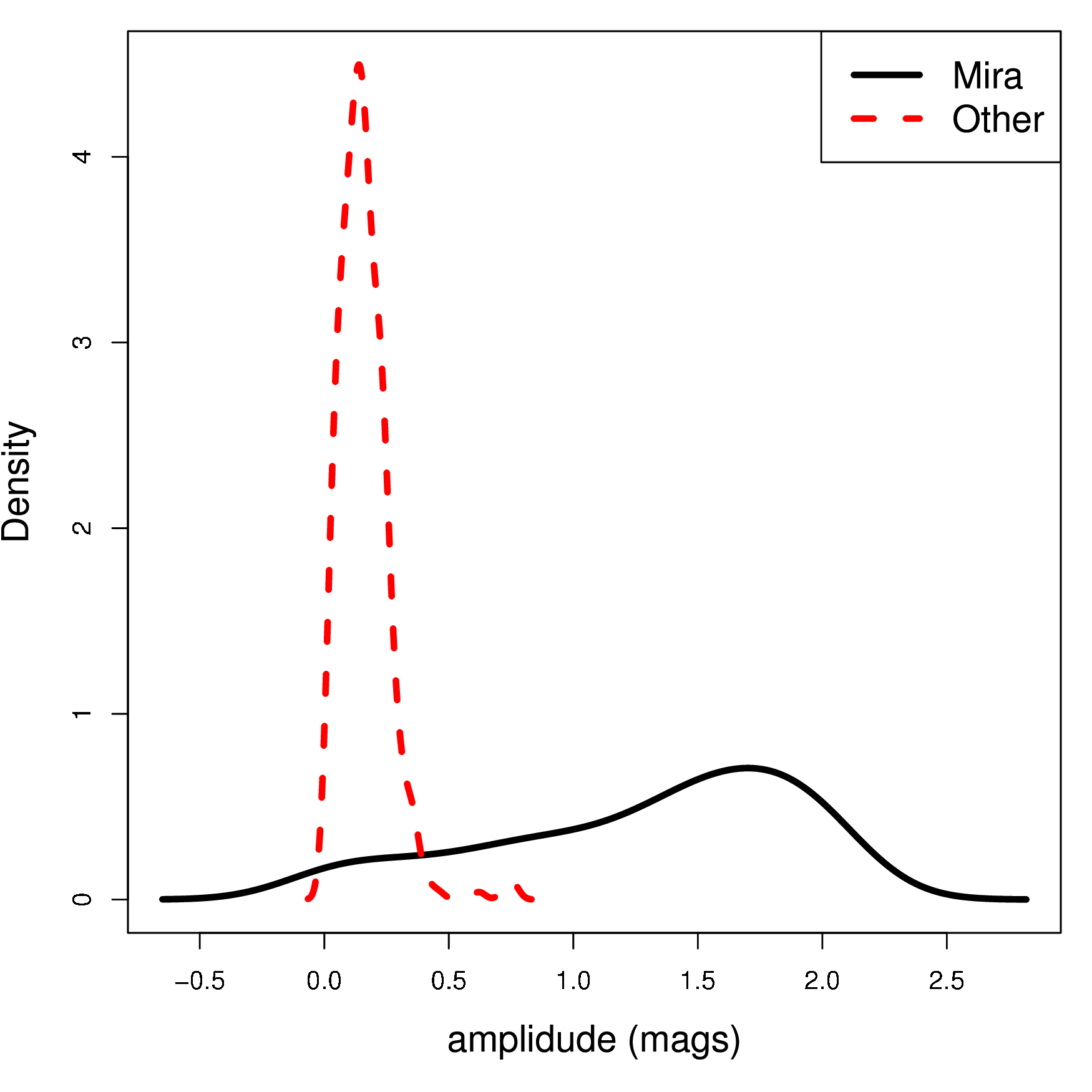}
\end{array}$
\end{center}
\caption{Amplitude feature distributions for Mira versus other classes for 10 flux measurements at (a) OGLE and (b) \hip{} cadence. The feature is very useful for separating Miras from non-Miras at the \hip{} cadence because of the irregular time sampling. Here we see how important it is for training and unlabeled data to have matching cadences, not just number of flux measurements.}
\label{fig:amplitude_simulated_hip_ogle}
\end{figure}

The difference in error rates between classifiers trained on data noisified to the cadence of the unlabeled data and those that are not suggests that at low number of flux measurements feature distributions are different for the OGLE and \hip{} cadences. To investigate this in Figure \ref{fig:amplitude_simulated_hip_ogle} we plot densities of amplitude for simulated light curves with 10 flux measurements at the OGLE and \hip{} cadences. To keep things simple we show two class densities -- Miras and not Miras. It is clear here that for the OGLE cadence amplitude is not a particularly useful feature for separating Miras from other sources whereas for the \hip{} cadence it is. Due to the regular sampling at one to two day intervals, 10 flux measurement OGLE curves have only captured part of a Mira period. Hence the amplitude of the source looks much smaller than it actually is. In contrast the large gaps between flux measurements in \hip{} cadences result in us observing a much larger piece of phase space and thus obtaining a better estimate of amplitude.

\subsection{Using \hip{} to Classify OGLE}
\label{sec:hip_to_classify_ogle}

Now that we have studied noisification in some controlled settings, we test the method on the original problem proposed in Section \ref{sec:intro}. Recall that we are classifying Miras, RR Lyrae AB, and Classical Cepheids Fundamental Mode using light curves from \hip{} as the training data and V-band OGLE as the unlabeled data. In Section \ref{sec:intro} we saw that training a classifier on the \hip{} curves and applying it directly to OGLE resulted in poor performance due, in part, to differences in number of flux measurements, cadence, and photometric error between the two data sets.

Table \ref{tab:training_vs_unlabeled} highlights some important differences between the \hip{} and V-band OGLE sources. See \cite{udalski2008photometry,soszynski2008optical,soszynski2009opticalRR,soszynski2009optical} for descriptions of OGLE III photometry and these three source classes.\footnote{These OGLE III sources are available here: \url{http://ogledb.astrouw.edu.pl/~ogle/CVS/}.} We use all OGLE III sources from the LMC belonging to the three classes of interest.

There are systematically fewer flux measurements in OGLE than in \hip{}. Unlike the previous example with I-band OGLE, the V-band OGLE curves here are fairly sparse. 25\% percent of the flux measurements are spaced 16 or more days apart. Perhaps the most striking difference between surveys is in the class proportions. RR Lyrae AB make up 26.6\% of light curves in \hip{} and 84.1\% of light curves in OGLE. This is most likely due to \hip{} magnitude limits which result in undersampling the intrinsically faint RR Lyrae AB relative to Mira and Classical Cepheids AB.

\begin{table}
{\small
\begin{center}
\begin{tabular}{cccccc}
\hline
\textbf{Survey} & \textbf{\# Sources} & \textbf{Class Probs.}$^a$ & \textbf{F / LC}$^b$ & 
\textbf{Time Diff$^c$} & \textbf{Error$^d$}\\
\hline
\hip{}$^e$ (training) & 357  & (0.45,0.27,0.28) & 81-119 & 0.01-0.25 & 0.015-0.034\\
OGLE (unlabeled) & 20605  & (0.09,0.84,0.07) & 36-74 & 5.1-16.0  & 0.022-0.050\\
\end{tabular}
\caption{Training and unlabeled set characteristics for example in Section \ref{sec:intro} and Subsection \ref{sec:hip_to_classify_ogle}.\label{tab:training_vs_unlabeled}}
\end{center}
{\tiny
$^a$ Class probs. is the class proportion of (Classical Cepheids F, RR Lyrae AB, Mira).\\
$^b$ F / LC is the first and third quartiles of flux measurements per light curve for training.\\
$^c$ Time Diff is the first and third quartiles of time differences in days between successive flux measurements.\\
$^d$ Error is the first and third quartiles of estimated photometric error in magnitude for all flux measurements. \\
$^e$ Light curves and classifications from \cite{richards2011machine}.
}}
\end{table}

To classify the OGLE sources, we noisify all the \hip{} light curves to OGLE cadence at 10 through 100 flux measurements in multiples of 10. We then construct classifiers on each of these sets, resulting in 10 noisified classifiers. Each OGLE light curve is classified using the classifier with the closest number of flux measurements. So for an unlabeled OGLE light curve with 27 flux measurements, we classify it using the noisified classifier constructed on the 30-flux measurement training set.

Table \ref{fig:naive} displays a confusion matrix for the classifier constructed on the unmodified \hip{} light curves when it is applied to the OGLE light curves. Table \ref{fig:hipno_adjust} shows the error rate using the noisification procedure. The overall error rate drops from 27\% to 7\% as a result of using noisification. This is driven by the drop in error rate for RR Lyrae AB (31\% error using unmodified classifier, 7\% after noisification) and the prevalence of RR Lyrae AB in OGLE. The error rate for Classical Cepheids F actually increases from 2\% to 10\% while for Miras it is roughly the same.

\begin{table}[ht]
\begin{center}
\begin{tabular}{ccccc|c}
&&\multicolumn{3}{c}{Predicted} \\ 
& & ClsC & Mira & RRLA & Err.Rate \\ 
  \cline{2-6}
&  ClsC & 1799 & 0 & 34 & 0.02 \\ 
True &  Mira & 58 & 1360 & 20 & 0.05 \\ 
&  RRLA & 5358 & 78 & 11898 & 0.31 \\ 
   \cline{2-6}
&Err.Rate & 0.75 & 0.05 & 0 & 0.27 \\ 
  \end{tabular}
\caption{Confusion matrix for classifier constructed on the unmodified Hipparcos light curves and applied to OGLE. Rows are true class and columns are predictions. The overall error rate is driven by the performance on the most abundant class, RR Lyrae AB.}
\label{fig:naive}
\end{center}
\end{table}
\begin{table}[ht]
\begin{center}
\begin{tabular}{ccccc|c}
&&\multicolumn{3}{c}{Predicted} \\ 
&  & ClsC & Mira & RRLA & Err.Rate \\ 
  \cline{2-6}
&ClsC & 1644 & 1 & 188 & 0.1 \\ 
True &  Mira & 18 & 1381 & 39 & 0.04 \\ 
&  RRLA & 1168 & 76 & 16090 & 0.07 \\ 
   \cline{2-6}
& Err.Rate & 0.42 & 0.05 & 0.01 & 0.07 \\ 
  \end{tabular}
\caption{Confusion matrix for classifier constructed on noisified Hipparcos light curves. Rows are true class and columns are predictions. The overall error rate has dropped to .07 from .26. This is due to better predicting RR Lyrae AB sources. The error rate on Classical Cepheids has actually increased.}
\label{fig:hipno_adjust}
\end{center}
\end{table}

\begin{figure}[ht]
\epsscale{0.29999999999999999}
\begin{center}
$\begin{array}{ccc}
\multicolumn{1}{l}{\mbox{(a)}} & \multicolumn{1}{l}{\mbox{(b)}} & \multicolumn{1}{l}{\mbox{(c)}} \\ \\ \\ [-.35in]
\plotone{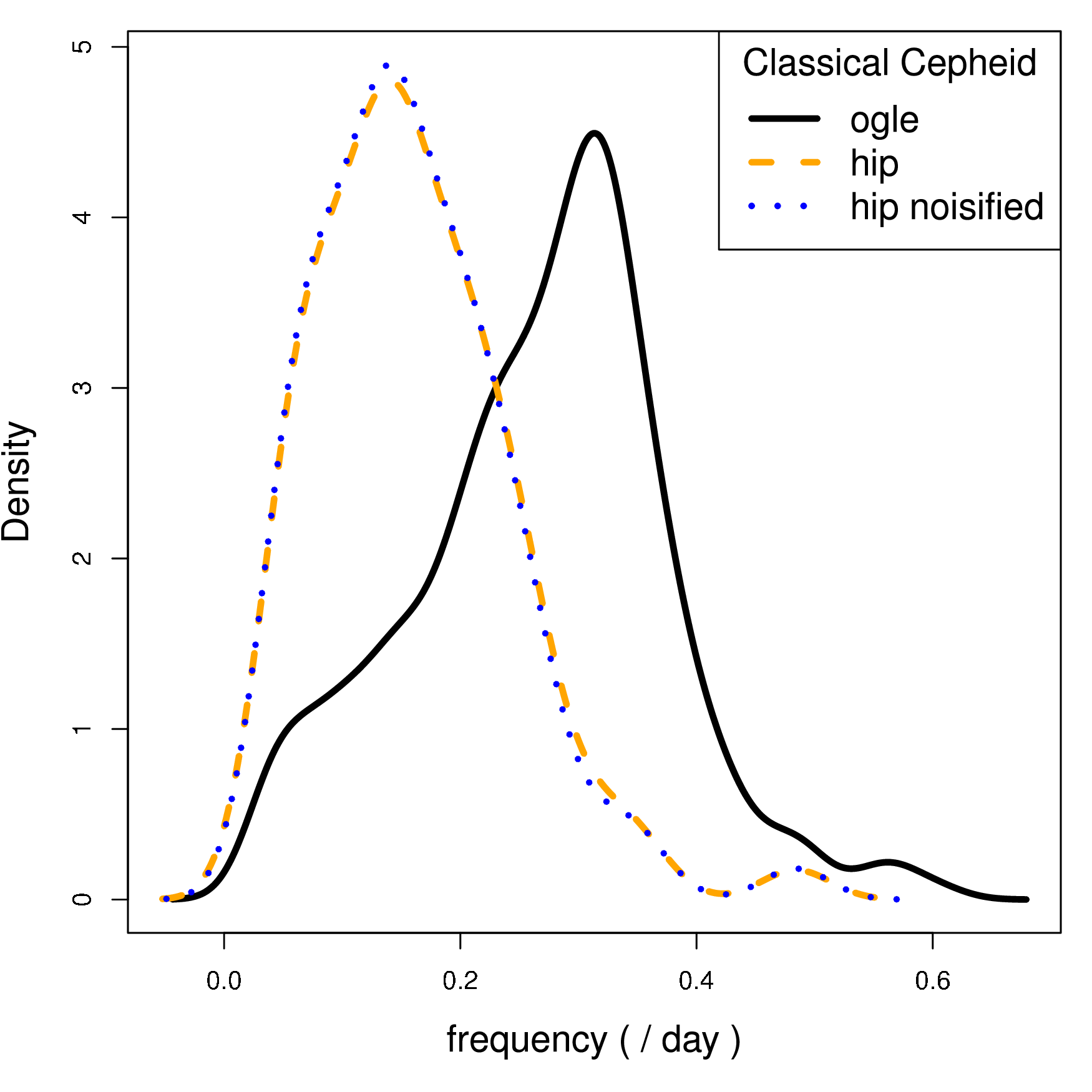} & 
\plotone{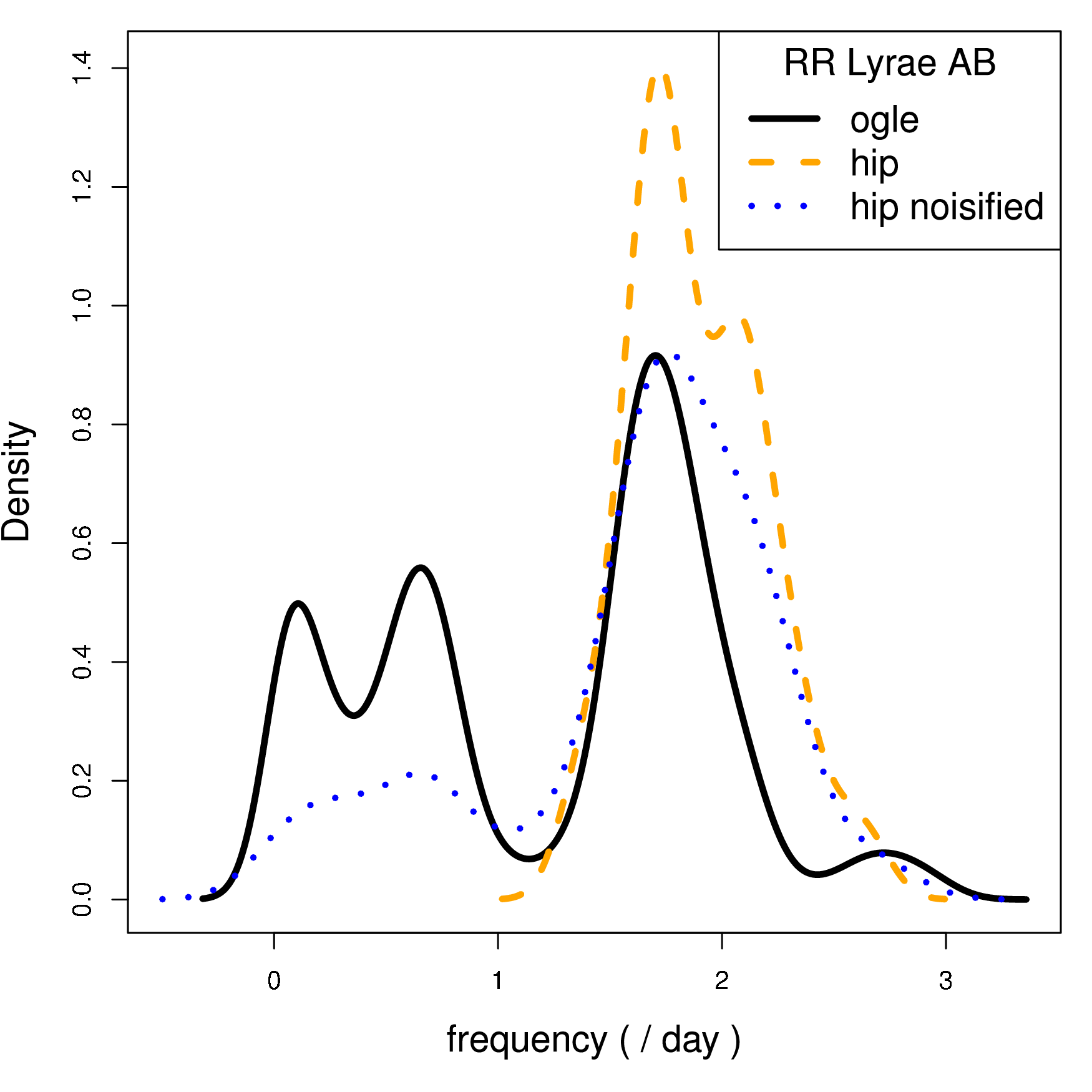} &
\plotone{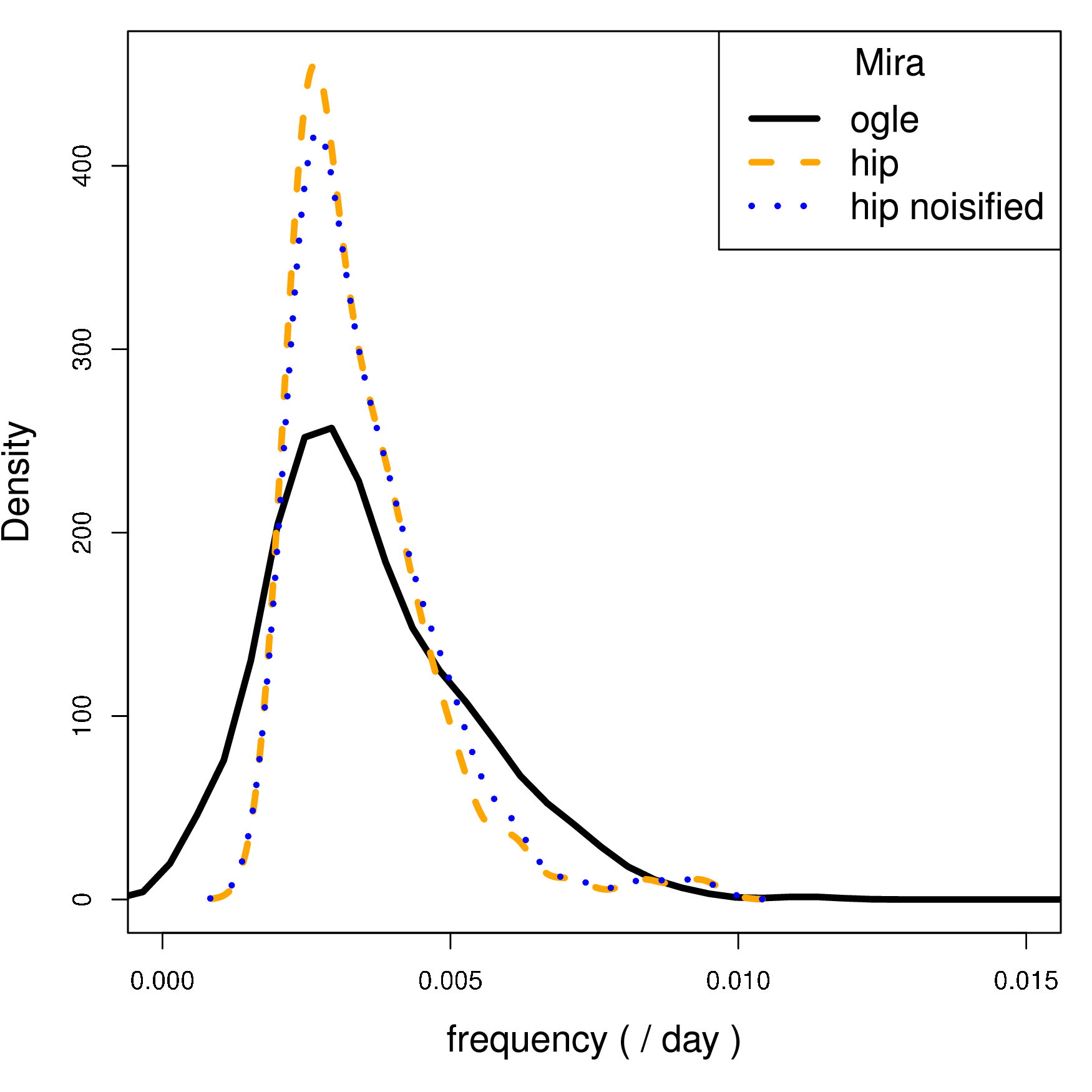}
\end{array}$
\end{center}
\caption{Density of frequency in OGLE light curves with 35--45 flux measurements (black solid), \hip{} before noisification (blue dots) and after noisification to 40 flux measurements (orange dashed) for (a) Classical Cepheids F, (b) RR Lyrae AB, and (c) Miras. Noisification of Cepheid periods does not match training and unlabeled densities because of survey differences not related to number of flux measurements, cadence or photometric error.}
\label{fig:cepheids_issue}
\end{figure}

Part of the reason why noisification increases the error rate for Classical Cepheids appears due to differences in distribution of frequency caused by Hipparcos magnitude limits. Figure \ref{fig:cepheids_issue} displays frequency density in \hip{}, 35-45 flux length OGLE, and \hip{} noisified to 40 flux for Cepheids (\ref{fig:cepheids_issue}a), RR Lyrae (\ref{fig:cepheids_issue}b), and Miras (\ref{fig:cepheids_issue}c). Noisification has not changed the density at all for the Cepheid sources (the blue and orange density almost exactly overlap) for the Cepheids. Visual inspection of OGLE periods revealed that they were correct. This suggests that the frequency distribution for Cepheids is fundamentally different in \hip{} and OGLE. This is likely due to magnitude limits in \hip{} and OGLE.

Lower frequency Cepheids are intrinsically brighter, so we can see them from further away. These low frequency Cepheids are over-represented in \hip{}. In contrast OGLE is closer to a random sample of Cepheids in the Large Magellanic Cloud (LMC). If it is there, we see it. Since this survey difference is not caused by number of flux measurements, cadence, or photometric error, the current implementation of noisification does not correct for it. Notice that in Figure \ref{fig:cepheids_issue} right plot, the noisification procedure has shifted the distribution of RR Lyrae frequencies in \hip{} to more closely match that in OGLE. Here much of the density mismatch was due to error in estimation of frequency due to having few flux measurements. Noisification helps us overcome this survey difference.

\begin{figure}[ht]
\epsscale{0.29999999999999999}
\begin{center}
$\begin{array}{ccc}
\multicolumn{1}{l}{\mbox{(a)}} & \multicolumn{1}{l}{\mbox{(b)}} & \multicolumn{1}{l}{\mbox{(c)}} \\ \\ \\ [-.35in]
\plotone{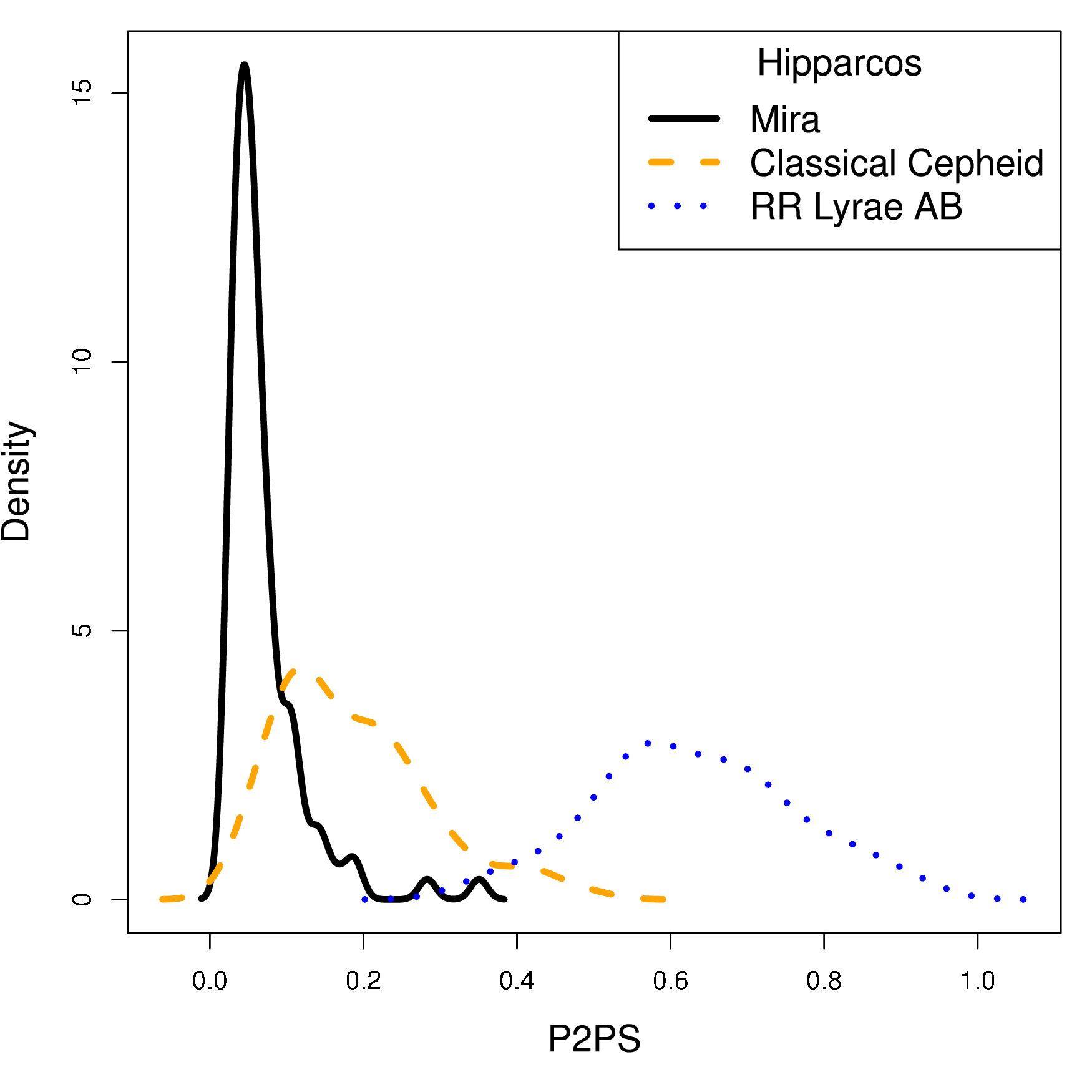} & 
\plotone{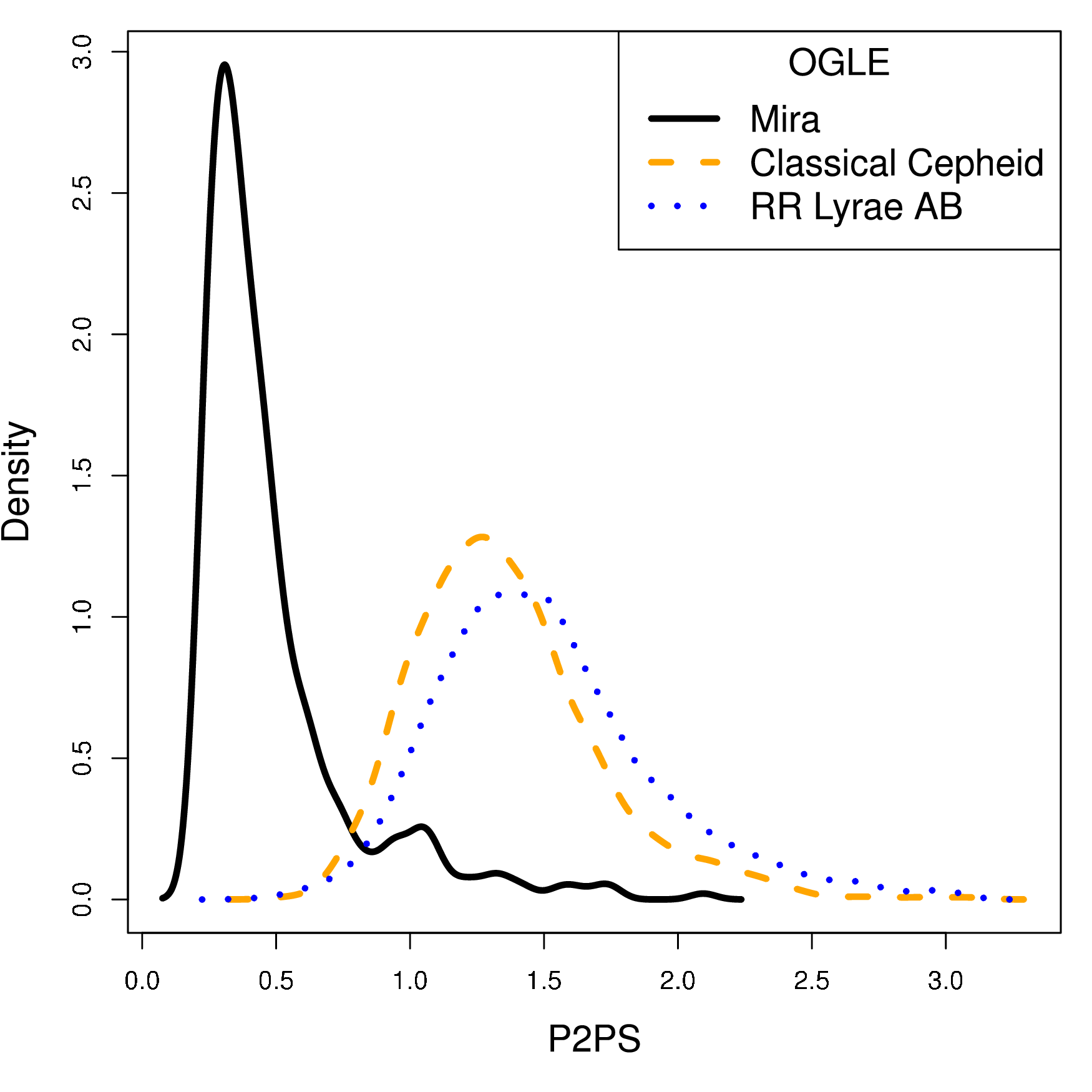} &
\plotone{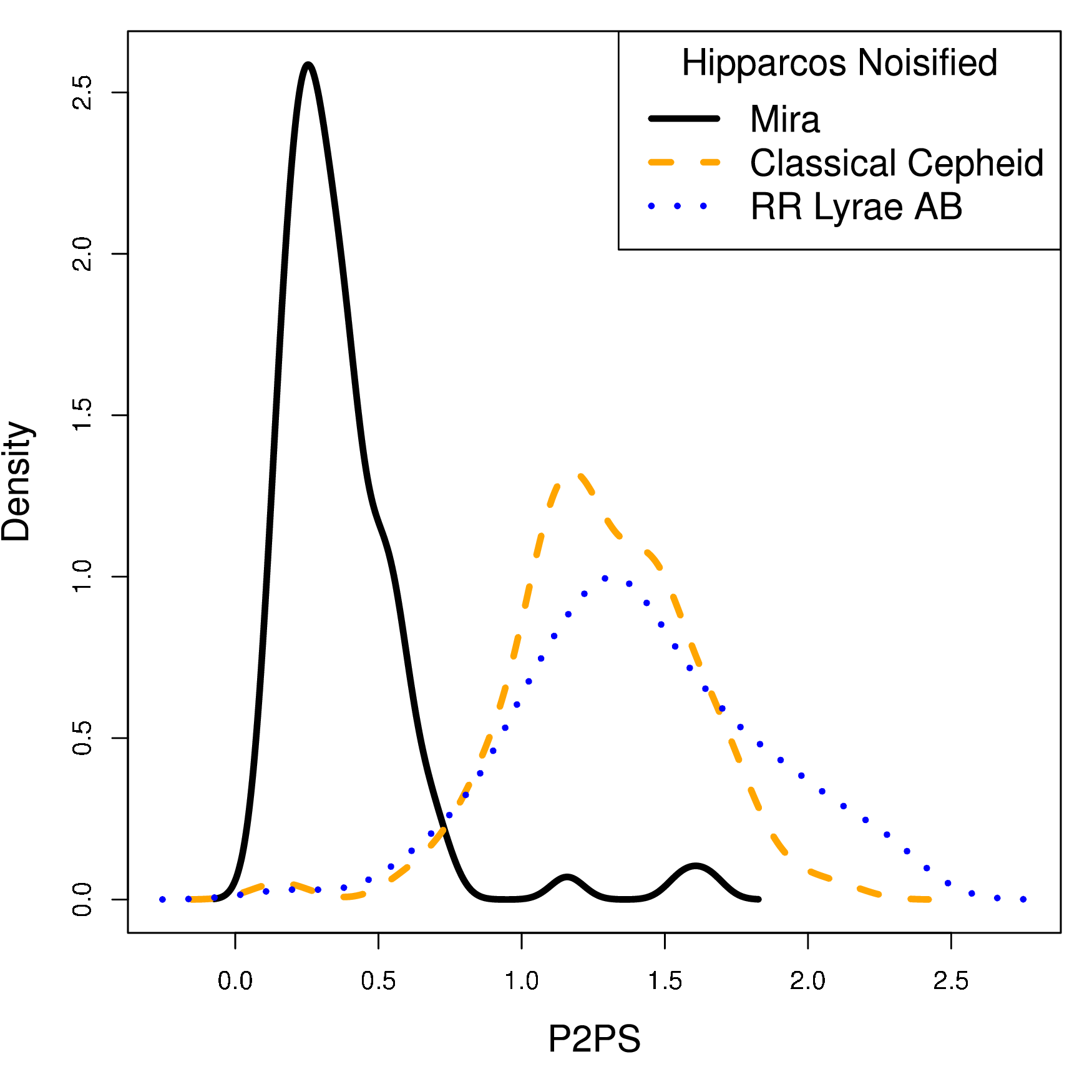}
\end{array}$
\end{center}
\caption{(a) P2PS in \hip{} un-noisified data. The feature appears useful for separating RR Lyrae from Miras and Classical Cepheids. (b) P2PS in OGLE for light curves with 35--45 flux measurements. Now Classical Cepheids have nearly the same density as RR Lyrae. A classifier trained on the un-noisified \hip{} light curves would not capture this relationship. (c) P2PS for \hip{} light curves noisified to OGLE cadence with 40 flux measurements. The densities now closely resemble the OGLE light curves.}
\label{fig:p2p_hip_ogle}
\end{figure}

Noisification is successful at matching other feature distributions. Figure \ref{fig:p2p_hip_ogle} displays the densities of P2PS for each sources class in \ref{fig:p2p_hip_ogle}a \hip{}, \ref{fig:p2p_hip_ogle}b OGLE, and \ref{fig:p2p_hip_ogle}c \hip{} noisified. There is a great deal of difference between \hip{} and OGLE densities. However the noisified \hip{} source densities appear to closely match the densities of OGLE.

\section{Conclusions}
\label{sec:conclusions}

We have highlighted how differences between training and unlabeled light curves induce different feature distributions. We then showed how these shifts in distribution can cause high error rates, even on problems where the unlabeled data is well separated in feature space. Common methods to evaluate classifier performance, such as cross--validation, do not detect these shifts in distribution and may give a false impression of classifier quality as they only reveal how well a classifier performs on data that is similar to the training set.

We developed a methodology, noisification, for overcoming differences between training and unlabeled data sets. As implemented in this study, noisification addresses differences due to the number of flux measurements, cadence, and photometric error. On several simulated and real--world examples, noisification greatly improved classifier performance. In the \hip{} training--OGLE unlabeled example, noisification reduced the misclassification rate by 20\%.

We hope these findings motivate practitioners to carefully consider differences between training and unlabeled data sets. In general, we recommend using training sets that match as closely as possible the unlabeled set of interest rather than training sets that are high signal--to--noise. As demonstrated in many examples, high signal--to--noise light curves often work poorly as training sets when the unlabeled light curves are of low quality. This is due to the classifier discovering class boundaries in feature space as they exist in the training set, not as they exist in the unlabeled set.

This study has made us skeptical of attempts to identify a single set of features that is generically sufficient for separating a set of classes of periodic variables. Useful features change depending on how sources are observed. The Random Forest importance plots (Figure \ref{fig:var_imp}) and the P2PS simulation (Subsection \ref{ss:cadence}) illustrate this. When implementing noisification, we recommend starting with large feature sets, even including features that are not useful for separating classes in the training data. These features may become useful for separating classes once the light curves have been noisified.

While we have studied noisification in the context of classification, it could also be applied to other problems. For example, novelty detection and unsupervised learning (clustering) methods are likely to work poorly when training and unlabeled data sets have systematic differences. Noisifying light curves offers a way to overcome these differences. 

Noisification may also be extended from what is implemented here to account for differences not related to number of flux measurements, cadence, and level of photometric error. For example, known censoring thresholds in the unlabeled data could be incorporated into the training data by removing, or marking as censored, flux measurements which would not have been observed in the unlabeled data set due to magnitude limits.

In the future, we will apply noisification to light curves from more surveys using larger, highly multi-class training sets. In parallel, we are developing a theoretical understanding of how noisification works and the problems for which it is most suitable. Of particular interest is how noisification performs when there are survey differences not addressed by noisification. This was the case with the Cepheid frequencies in the three--class \hip{}--OGLE problem.

Upcoming surveys pose a challenge based in their size \textit{and} their novelty. Not only will Gaia and LSST detect orders of magnitude more periodic variables than previous surveys, the sources they collect will have different properties than any training data we currently have. Noisification offers the potential to bridge some of these differences, enabling us to optimize scientific discovery.

The authors would like to thank Laurent Eyer and Dan Starr for helpful comments and criticisms. The authors would like to acknowledge the generous support of a Cyber-Enabled Discovery and Innovation (CDI) grant (No. 0941742) from the National Science Foundation. This work was performed in the CDI-sponsored Center for Time Domain Informatics (\url{http://cftd.info}).

\appendix
\section{Description of Features}
\label{app:new_features}

We used 62 features in this work. Fifty of these features came from Tables 4 and 5 in \cite{richards2011machine}. We did not use the features \textbf{pair\_slope\_trend}, \textbf{max\_slope}, or \textbf{linear\_trend} from these tables. We used 12 additional features. Five are from \cite{dubath2011random}.\footnote{\textbf{scatter\_res\_raw}, \textbf{medperc90\_2p\_p}, \textbf{p2p\_scatter\_2praw}, \textbf{P2PS} (named P2p\_scatter in \cite{dubath2011random}), and \textbf{p2p\_scatter\_pfold\_over\_mad}} The remaining seven are:

\begin{enumerate}
\item \textbf{fold2P\_slope\_10percentile} 10th percentile of slopes between adjacent flux measurements after the light curve has been folded on twice the estimated period
\item \textbf{fold2P\_slope\_90percentile} 90th percentile of slopes between adjacent flux measurements after the light curve has been folded on twice the estimated period
\item \textbf{freq\_frequency\_ratio\_21} ratio of the second to first frequency determined by lomb-scargle ($\frac{f_2}{f_1}$ from Table 4 in \cite{richards2011machine})
\item \textbf{freq\_frequency\_ratio\_31} ratio of the third to first frequency determined by lomb-scargle ($\frac{f_3}{f_1}$ from Table 4 in \cite{richards2011machine})
\item \textbf{freq\_amplitude\_ratio\_21} ratio of amplitude for frequency 2 to amplitude for frequency 1 ($\frac{A_{2,1}}{A_{1,1}}$ from Table 4 in \cite{richards2011machine})
\item \textbf{freq\_amplitude\_ratio\_31} ratio of amplitude for frequency 3 to amplitude for frequency 1 ($\frac{A_{3,1}}{A_{1,1}}$ from Table 4 in \cite{richards2011machine})
\item \textbf{p2p\_ssqr\_diff\_over\_var}\footnote{From \cite{kim2011qso}} the sum of squared differences in successive flux measurements divided by the variance of the flux measurements
\end{enumerate}

\bibliographystyle{apj}
\bibliography{transientsbib}

\end{document}